\documentclass[useAMS,usenatbib]{mn2e}

\usepackage{times}  
\usepackage{listings}
\usepackage{graphics}
\usepackage{subfigure}
\usepackage{graphicx}
\usepackage{graphicx,xspace}
\usepackage{amssymb}
\usepackage{amsmath}
\usepackage{aas_macros}
\usepackage{lscape}
\usepackage{rotating}
\usepackage{placeins}
\usepackage{natbib}
\usepackage{color}
     
\newcommand{\galform}{{\sc{galform}}\xspace}

\newcommand{\eagle}{{\sc{eagle}}\xspace}
\newcommand{\gadget}{{\sc{gadget-3}}\xspace}

\newcommand{\subfind}{{\sc{subfind}}\xspace}

\setcitestyle{notesep={ },round,aysep={},yysep={}}

\title[Inflows and wind recycling in EAGLE]
{Galactic inflow and wind recycling rates in the EAGLE simulations}

\author[P. D. Mitchell et al.]{
\newauthor Peter D. Mitchell\thanks{\rm E-mail: mitchell@strw.leidenuniv.nl}$^{1}$,
Joop Schaye$^{1}$
and Richard G. Bower$^{2}$
\\
$^{1}$Leiden Observatory, Leiden University, P.O. Box 9513, 2300 RA Leiden, the Netherlands\\
$^{2}$Institute for Computational Cosmology, Department of Physics, Durham University, South Road, Durham, DH1 3LE, UK\\
}

\begin{document}
\date{\today}
\pagerange{\pageref{firstpage}--\pageref{lastpage}} \pubyear{2018}
\maketitle
\label{firstpage}

\begin{abstract}
The role of galactic wind recycling represents one of the largest unknowns
in galaxy evolution, as any contribution of recycling to galaxy growth
is largely degenerate with the inflow rates of first-time infalling
material, and the rates with which outflowing gas and metals are driven from galaxies.
We present measurements of the efficiency of wind recycling from the 
\eagle cosmological simulation project, leveraging the statistical
power of large-volume simulations that reproduce a realistic galaxy population.
We study wind recycling at the halo scale, i.e. gas that has been ejected beyond the halo virial radius, and 
at the galaxy scale, i.e. gas that has been ejected from the ISM to 
at least $\approx 10 \, \%$ of the virial radius (thus excluding smaller-scale galactic fountains).
Galaxy-scale wind recycling is generally inefficient, with a characteristic return
timescale that is comparable or longer than a Hubble time, and with
an efficiency that clearly peaks at the characteristic halo mass of $M_{200} = 10^{12} \, \mathrm{M_\odot}$.
Correspondingly, the majority of gas being accreted onto galaxies
in \eagle is infalling for the first time. 
Recycling is more efficient at the halo scale, with values
that differ by orders of magnitude from those assumed by semi-analytic
galaxy formation models. Differences in the efficiency of wind recycling
with other hydrodynamical simulations
are currently difficult to assess, but are likely smaller. 
We are able to show that the fractional contribution of wind recycling to galaxy growth is smaller
in \eagle than in some other simulations.
In addition to measurements of wind recycling, we study
the efficiency with which first-time infalling material is accreted
through the virial radius, and also the efficiency with which this
material reaches the ISM.
We find that cumulative first-time gas accretion rates at the virial radius
are reduced relative to the expectation from dark matter accretion for haloes
with mass, $M_{200} < 10^{12} \, \mathrm{M_\odot}$, indicating efficient
preventative feedback on halo scales.
\end{abstract}


\section{Introduction}

In the modern cosmological paradigm, galaxies are thought to form within dark matter haloes, 
which represent collapsed density fluctuations that grew from a near-uniform density
field via gravitational instability. Dark matter haloes grow gradually by the
accretion of smaller haloes, and baryonic accretion onto haloes is expected to trace
this process, with half of the current stellar mass density of the Universe
having formed after $z \approx 1.3$ \cite[][]{Madau14}. In this picture, actively
star-forming galaxies continually accrete gas from their wider environments,
and this in turn helps to explain the observed chemical abundances of stars
\cite[e.g.][]{Larson72}, and the relatively short inferred gas depletion 
timescales of star-forming galaxies \cite[e.g.][]{Scoville17}.

While there is strong theoretical and indirect observational evidence
for sustained gas accretion onto the interstellar medium (ISM) of galaxies, direct measurements of
gaseous inflow rates have remained inaccessible, owing primarily to the tenuous
low-density nature of extra-galactic gas, and to the weak expected
kinematic signature (relative for example to the very strong kinematic
signature of feedback-driven galactic outflows).
Various observations that trace inflowing gas have been reported
however, both for the Milky Way and for extra-galactic sources
\cite[e.g.][]{Rubin12,Fox14,Turner17,Bish19,Roberts-Borsani19,Zabl19}.

With a paucity of strong observational constraints, cosmological 
simulations have been used extensively as an
alternative way to study gas accretion onto haloes and galaxies 
\cite[e.g.][]{Keres05,FaucherGiguere11,VanDeVoort11,Nelson13,RamanoDiaz17,Vandevoort17,Correa18a,Correa18b}.
Simulations have predicted, for example, the presence of filamentary accretion
streams, reflecting the larger scale filamentary structure of
the cosmic web \cite[e.g.][]{Katz03,Dekel06}. Simulation
predictions for inflows are expected to be strongly model dependent
however, since it has been demonstrated that feedback processes (the
implementation of which remains highly uncertain in simulations) 
modulate gaseous inflow rates, either by reducing the rate
of first-time gaseous infall \cite[e.g.][]{Nelson15}, or by the
recycling of previously ejected wind material \cite[e.g.][]{Oppenheimer10}.

The spatial and halo mass scales for which these processes
play a significant role have been studied in a variety
of simulations. Broadly speaking, feedback processes are
expected to strongly modulate the accretion rates of
gas onto the ISM of galaxies \cite[e.g.][]{Oppenheimer10,FaucherGiguere11,VanDeVoort11,Nelson15,AnglesAlcazar17,Correa18b},
and to shape the content of the circum-galactic medium
\cite[e.g.][]{Hafen19}.
The effect of feedback processes on inflows at the scale of the
halo virial radius is less clear, although effects
are often reported for haloes with masses 
$M_{200} < 10^{12} \, \mathrm{M_\odot}$ 
\cite[][]{VanDeVoort11,FaucherGiguere11, Christensen16, Tollet19}.
The role of wind recycling is also debated, with studies
reporting that between $50 \, \%$ \cite[][]{Ubler14} 
and $90 \, \%$ \cite[][]{Grand19} of gaseous inflow onto galaxies is recycled
for different cosmological simulations 
\cite[see][for a recent review]{Vandevoort17b}. 
Furthermore, recent studies have highlighted the potential
importance of the transfer of gas between galaxies \cite[][]{AnglesAlcazar17},
and haloes \cite[][]{Borrow20}, due to feedback related processes.

The potential role of wind recycling has also been explored in more
idealised analytic and semi-analytical models of galaxy formation.
In particular, authors have highlighted how a strong dependence
of the efficiency of wind recycling with halo mass can help reconcile
models with the observed evolution of the galaxy stellar mass function
\cite[][]{Henriques13,Hirschmann16}, and that a strongly time-evolving
recycling efficiency can explain the observed evolution
of galaxy specific star formation rates \cite[][]{Mitchell14}, which is generally not
reproduced in models and simulations \cite[e.g.][]{Daddi07,Mitchell14,Kaviraj17,Pillepich18}.
Generally speaking, these studies demonstrate that 
gas recycling (if it proceeds over timescales that are comparable or longer than 
the other timescales that govern galaxy growth, and with
strong mass and/or redshift dependence) is a highly promising mechanism
for decoupling the growth of
galaxies from the growth of their host dark matter haloes, which
is required to explain various observational trends.

Recent years have seen the development of cosmological simulations
that produce a relatively realistic population of galaxies when compared
to current observational constraints, and that simulate the galaxy
population over a representative volume 
\cite[e.g.][]{Vogelsberger14,Schaye15}.
The statistical sample sizes afforded by such simulations greatly facilitate
the study of correlations between gaseous inflows and other galaxy
properties \cite[such as radial metallicity gradients,][]{Collacchioni19}, and to study the role of environment
on accretion \cite[][]{Vandevoort17}. The realism of these simulations affords
additional confidence to the results, in contrast
to the older simulations that did not reproduce the observed
galaxy stellar mass function. As an example, \cite{Oppenheimer10} 
find that wind recycling dominates gas accretion onto massive
galaxies, but their simulations do not include AGN feedback, and so
greatly over-predict the abundance of massive galaxies.

As one of the current state-of-the-art modern large-volume cosmological simulations,
the \eagle simulation project simulates the formation and evolution 
of galaxies within the $\Lambda$ Cold Dark Matter model, integrating 
periodic cubic boxes (up to $100^3 \, \mathrm{Mpc^{3}}$ in volume) 
down to $z=0$ \cite[][]{Schaye15,Crain15}. With a fiducial baryonic 
particle mass of $1.81 \times 10^6 \, \mathrm{M_\odot}$,
\eagle resolves galaxies (with at least $100$ stellar particles) over
roughly five orders of magnitude in halo mass ($10^{11} < M_{200} < 10^{14}$).
\eagle has been used to study gas accretion onto haloes and galaxies,
with individual studies focussing on the dichotomy between cold and
hot accretion \cite[][]{Correa18a}, the impact of changing feedback
models on gas accretion \cite[][]{Correa18b}, the impact of environment
\cite[][]{Vandevoort17}, the angular momentum content of cooling
coronal gas \cite[][]{Stevens17}, and the connection between accretion
and radial metallicity gradients in galaxies \cite[][]{Collacchioni19}.

In this study, we extend these analyses by using \eagle to explicitly track the
gas that is ejected from galaxies and haloes, which enables quantitative
measurements of the efficiency and role of recycled accretion, as
well as the study of gas that is transferred between independent
galaxies and haloes. This study also follows from \cite{Mitchell20},
in which we present measurements of outflows on galaxy and halo
scales. In future work, we then intend to combine these measurements 
together, in order to explicitly study how 
the mass and redshift scalings for first-time inflows, outflows,
and wind recycling act in conjunction to explain the origin 
and evolution of the scaling relations between galaxy stellar mass
and halo mass, and between galaxy star formation rate and stellar mass.

Relative to other studies of inflows and recycling in cosmological
simulations \cite[e.g.][]{AnglesAlcazar17,Grand19}, we take care to 
measure quantities that can be robustly mapped onto
a simplified description of galaxy formation as a network
of ordinary differential equations. Most pertinently, we track the
evolution of gas that is accreted onto galaxies or haloes
at any time during their evolution, rather than only the subset
of stars and gas that is located within a galaxy at some final
redshift of selection. This means we can assess
the characteristic timescale for all ejected gas to return,
rather than for only the subset of gas that has returned by a given 
redshift. We also attempt to (as far as is reasonably possible)
present a robust comparison of how wind recycling proceeds
between various recent simulations and models from the
literature, and in doing so identify areas of consensus
(or tension) in the current theoretical picture.

The layout of this study is as follows: we present details of
the \eagle simulations and our methodology 
in Section~\ref{methods_section}, we present our main results 
in Section~\ref{results_section}, a comparison with other recent theoretical studies
from the literature is presented in Section~\ref{literature_section}, and we
summarise our results and conclusions in Section~\ref{summary_section}.

\section{Methods}
\label{methods_section}

\subsection{Simulations and subgrid physics}

We utilise the \eagle suite of cosmological hydrodynamical simulations \cite[][]{Schaye15},
which have been publicly released \cite[][]{McAlpine16}.
\eagle simulates cubic periodic boxes of representative volumes with full gravity and 
smoothed particle hydrodynamics (SPH) using a modified version of the \gadget code
\cite[last described in][]{Springel05b}, 
and employs uniform resolution throughout each simulation. 
A $\Lambda$CDM cosmological model is assumed, with parameters set following \cite{Planck14}.
``Subgrid'' physics are implemented to account for relevant physical processes
that are not resolved (e.g. star formation), and radiative cooling and heating
are modelled assuming a uniform ultraviolet radiation background, assuming
the gas to be optically thin and in ionization equilibrium.
Ionization modelling is performed for $11$ elements. 

The simulation suite includes a \emph{Reference} set of parameters for the subgrid 
physics model, calibrated at a fiducial numerical resolution to reproduce the following observational 
diagnostics of the $z \approx 0$ galaxy population: the galaxy stellar mass function, 
the relationship between galaxy size and stellar mass, and the relationship between 
supermassive black hole (SMBH) mass and galaxy stellar mass.
The largest \eagle simulation simulates a $(100 \, \mathrm{Mpc})^3$ volume
with $2 \times 1504^3$ particles, with a fiducial particle mass of 
$1.8 \times 10^6 \, \mathrm{M_\odot}$ for gas and $9.7 \times 10^6 \, \mathrm{M_\odot}$
for dark matter.
Unless otherwise stated, all of the \eagle measurements presented in this article are 
taken from this simulation. 
The suite also contains simulations run with variations of the Reference model
parameters, including a simulation named \emph{Recal}, for which the model
parameters were (re)calibrated against the same observational constraints,
but at eight times higher numerical mass resolution than that of the Reference model.

Star particles are allowed to form from gas particles that first pass the metallicity-dependent
density threshold for the transition from the warm, atomic to the cold, molecular ISM, 
as derived by \cite{Schaye04}:

\begin{equation}
n_{\mathrm{H}}^{\star} = \mathrm{min}\left(0.1 \left(\frac{Z}{0.002}\right)^{-0.64},10 \right) \, \mathrm{cm^{-3}},
\label{eqn_sf_thr}
\end{equation}

\noindent where $Z$ is the gas metallicity. In addition, star formation is restricted to gas particles with temperature
within $0.5 \, \mathrm{dex}$ from a temperature floor, $T_{\mathrm{eos}}$, which corresponds
to an imposed equation of state $P_{\mathrm{eos}} \propto \rho^{4/3}$, normalised
to a temperature of $T = 8 \times 10^3 \, \mathrm{K}$ at a hydrogen number density
of $0.1 \, \mathrm{cm^{-3}}$ \cite[][]{Schaye08}. Gas particles are artificially pressurised up
to this floor, such that in practice the ISM of galaxies
is stabilised in a warm phase, preventing radiative cooling from leading to
runaway fragmentation on Jeans scales that are unresolved in the simulation. 

Eligible gas particles are turned into stars stochastically, with the average rate
given by 

\begin{equation}
\psi = m_{\mathrm{gas}} \, A (1 \mathrm{M_\odot} \mathrm{pc}^{-2})^{-n} \, \left(\frac{\gamma}{G} f_{\mathrm{g}} P\right)^{(n-1\
)/2},
\label{sfr_eqn_eagle}
\end{equation}

\noindent where $P$ is the local gas pressure, $m_{\mathrm{gas}}$ is the gas particle mass,
$\gamma=5/3$ is the ratio of specific heats, $G$ is the gravitational constant, $f_{\mathrm{g}}$ is the gas mass
fraction (set to unity). As described in \cite{Schaye08}, this corresponds to
a Kennicutt-Schmidt law for a gas disk in vertical hydrostatic equilibrium,
with the dependent variable transformed from gas surface density to pressure.
Following observational constraints on the observed Kennicutt-Schmidt law, 
$A$ and $n$ are set to $A = 1.515 \times 10^{-4} \, \mathrm{M_\odot yr^{-1} kpc^{-2}}$
and $n=1.4$ \cite[][]{Kennicutt98}.

A simple stellar feedback model is implemented in \eagle that conceptually 
accounts for the combined effects of energy injected into the ISM by radiation 
and stellar winds from young stars, as well as supernova explosions.
Thermal energy is injected stochastically by a fixed temperature difference
of $\Delta T  = 10^{7.5} \, \mathrm{K}$, with a high value above the peak
of the radiative cooling curve chosen to mitigate the effects of spurious 
radiative losses that are expected to occur if the injected energy were to instead be
spread more uniformly in a poorly resolved artificially pressurised warm medium \cite[][]{DallaVecchia12}.
No kinetic energy or momentum is injected directly by the subgrid model.
Stellar feedback energy is injected when stars reach an age of $30 \, \mathrm{Myr}$,
at a rate set such that the average energy injected is 
$f_{\mathrm{th}} \times \, 8.73 \times 10^{15} \, \mathrm{erg \, g^{-1}}$ of
stellar mass formed, where $f_{\mathrm{th}}$ is a model parameter. For $f_{\mathrm{th}} = 1$,
the expectation value for the number of feedback events per particle is of
order unity, and the injected energy per unit stellar mass corresponds to that of a simple stellar
population with a Chabrier initial mass function, assuming that $6-100 \, \mathrm{M_\odot}$ stars explode as supernovae,
and that each supernova injects $10^{51} \, \mathrm{erg}$ of energy.

In practice, it was found that while adopting $f_{\mathrm{th}}=1$ reproduced
the observed galaxy stellar mass function reasonably well, it was necessary
to inject extra energy at the high densities for which numerical overcooling
is expected in order to also reproduce the 
observed galaxy size-stellar mass relation \cite[][]{Schaye15,Crain15}.
The following form was adopted

\begin{equation}
f_{\mathrm{th}} = f_{\mathrm{th,min}} + \frac{f_{\mathrm{th,max}} - f_{\mathrm{th,min}}}{1+\left(\frac{Z}{0.1 \mathrm{Z_\odot}}\right)^{n_{\mathrm{Z}}} \left(\frac{n_{\mathrm{H,birth}}}{n_{\mathrm{H,0}}}\right)^{-n_{\mathrm{n}}}},
\label{Eagle_eff}
\end{equation}

\noindent where $f_{\mathrm{th,min}}$ and $f_{\mathrm{th,max}}$ are model parameters that are the asymptotic values
of a sigmoid function in metallicity, with a transition scale at a characteristic
metallicity, $0.1 \mathrm{Z_\odot}$ \cite[above which radiative losses are expected to increase due to metal cooling,][]{Wiersma09b},
and with a width controlled by $n_{\mathrm{Z}}$.
The two asymptotes, $f_{\mathrm{th,min}}$ and $f_{\mathrm{th,max}}$,
are set to $0.3$ and $3$ respectively, such that between $0.3$ and $3$ times the
canonical supernova energy is injected. 
The dependence on local gas density is controlled by model parameters,
$n_{\mathrm{H,0}}$, and $n_{\mathrm{n}}$. For the Reference model, $n_{\mathrm{Z}}$ and $n_{\mathrm{n}}$
are both set to $2/\ln(10)$, and $n_{\mathrm{H,0}}$ is set to 
$1.46 \, \mathrm{cm^{-3}}$.\footnote{Note that the original value of 
$n_{\mathrm{H,0}} = 0.67 \, \mathrm{cm^{-3}}$ quoted in \cite{Schaye15} is
incorrect. This error does not however affect any of the quoted values of
$f_{\mathrm{th}}$ in that paper, with the mean and median values of $f_{\mathrm{th}}$
across the reference $(100 \, \mathrm{Mpc})^3$ simulation being $1.06$ and $0.7$ respectively.}

Supermassive black hole (SMBH) seeds are inserted into haloes with mass $> 10^{10} \, \mathrm{M_\odot} /h$, as
identified on the fly by a friend-of-friends (FoF) algorithm, using a linking length 
set equal to $0.2$ times the average inter-particle separation. SMBH particles then
grow by merging with other black holes, or by accreting gas particles at a rate
given by a version of the Bondi accretion, modified such that accretion is
reduced if the surrounding gas is rotating rapidly relative to the local sound speed
\cite[][]{Rosas15,Schaye15}.

Similar to stellar feedback, feedback from accreting SMBH particles is implemented
by stochastically heating neighbouring gas particles by a fixed temperature
jump \cite[][]{Booth09}, in this case set to $10^{8.5} \, \mathrm{K}$ for the Reference model.
Energy is injected on average at a rate given by 

\begin{equation}
\dot{E}_{\mathrm{AGN}} = \epsilon_{\mathrm{f}} \epsilon_{\mathrm{r}} \dot{m}_{\mathrm{acc}} c^2,
\label{Eagle_agn_fb}
\end{equation}

\noindent where $\dot{m}_{\mathrm{acc}}$ is the gas mass accretion rate onto the SMBH,
$c$ is the speed of light, $\epsilon_{\mathrm{r}}$ is the fraction of the
accreted rest mass energy which is radiated (set to $0.1$), and $\epsilon_{\mathrm{f}}$ is a model
parameter which sets the fraction of the radiated energy that couples to
the ISM (set to $0.15$).

\subsection{Subhalo identification and merger trees}

Haloes are identified in the simulations first using a FoF algorithm, with a
linking length set to $0.2$ times the average inter-particle separation.
Haloes are then divided into subhaloes using the \subfind algorithm
\cite[][]{Springel01,Dolag09}. Subhalo centers are set to the location
of the particle with the lowest value of the gravitational potential.
The subhalo within each FoF group with the lowest potential value is
considered to be the central subhalo, and other subhaloes are referred
to as satellites. For central subhaloes, we manually associate all
particles within $R_{\mathrm{vir}} = R_{\mathrm{200, crit}}$ to 
the subhalo for the purpose of computing accretion rates, etc, where
$R_{\mathrm{200, crit}}$ is the radius enclosing a mean spherical overdensity
equal to $200$ times the critical density of the Universe.
Accordingly, we quote halo masses as $M_{200}$, the mass contained
within this radius.

We use merger trees constructed according to the algorithm described
in detail in \cite{Jiang14}, with some additional post-processing
steps that are described in \cite{Mitchell20}. The selection
of the main progenitor of each subhalo is based on bijective 
matching between the $N_{\mathrm{link}}$ most-bound particles 
in each progenitor with 
those of the descendant, with $10 \leq N_{\mathrm{link}} \leq 100$,
depending on the total number of particles in each subhalo. 
This is then used to define the accretion rates of gas and dark matter that are associated
with halo or galaxy merging events.

\subsection{Measuring inflow rates with particle tracking}
\label{tracking_sec}

We measure inflow rates by tracking particles between consecutive simulation 
snapshots, exploiting the Lagrangian nature of the underlying SPH hydrodynamical
scheme. We choose to measure inflow rates at two scales: the first being
gas accretion onto haloes, and the second being gas accretion onto
the ISM of galaxies. Inflow onto haloes is measured by identifying particles that cross
the virial radius, and inflow onto galaxies involves identifying particles
that join the ISM. Note that the inflow rates quoted in this study
include only particles that join galaxies or haloes, and so do not represent
the \emph{net} inflow (inflow minus outflow) onto the system\footnote{Authors 
interested in the net inflow rates in \eagle can obtain
them by combining the results presented here with the outflow rates
presented in \cite{Mitchell20}.}.

We define the ISM as in \cite{Mitchell20}, including
particles that are star forming, meaning they are both within $0.5 \, \mathrm{dex}$ in
temperature of the density-dependent temperature floor corresponding to the imposed equation of state, 
and that they also pass the metallicity-dependent density threshold given by 
Eqn.~\ref{eqn_sf_thr}. We also include in the ISM any non-star-forming particles that are
still within $0.5 \, \mathrm{dex}$ of the temperature floor, and with
density $n_{\mathrm{H}} > 0.01 \, \mathrm{cm^{-3}}$, approximately
mimicking a selection of neutral atomic hydrogen.

We measure whether inflowing particles are being recycled onto galaxies (or haloes)
by first establishing if particles are ejected from the ISM of galaxies
(or through the halo virial radius), following the procedure introduced
and motivated by \cite{Mitchell20}. We impose a time-integrated radial velocity
cut to select particles that are genuinely outflowing from the galaxy or
from the halo (the instantaneous radial velocity is an unreliable predictor 
of whether particles will move outwards over a finite distance).
Particles that fail this cut are not included in any later inflow measurement
(neither first-time nor recycled). 
As a further detail, particles that fail the cut are given
the opportunity to pass the cut (and so join the outflow) at
later snapshots, until they have either rejoined the ISM (or halo)
or until three halo dynamical times have passed.

We use a fiducial time-integrated velocity
cut of $\frac{\Delta r_{21}}{\Delta t_{21}} > 0.25 \, V_{\mathrm{max}}$,
where $\Delta r_{21}$ is the radial distance moved between snapshots
$1$ and $2$ by a particle that is first recorded as having left
the ISM (or halo) at snapshot $1$, and $V_{\mathrm{max}}$ is the
subhalo maximum circular velocity. The time interval ($\Delta t_{21}$) between snapshots
$1$ and $2$ is held (as near as possible) constant to one quarter of a halo
dynamical time, which mitigates the implicit dependence of outflow selection 
on the underlying snapshot spacing. In practice, the selection corresponds
to a minimum radial displacement of 
$\approx 15 \, \mathrm{kpc}$ for a
halo mass of $10^{12} \, \mathrm{M_\odot}$ at $z=0$ (i.e. $7 \, \%$ of the halo
virial radius, which is the case almost independently of halo mass, but changes
to a slightly larger fraction of $R_{200}$ at higher redshifts). The impact of changing this velocity
cut by a factor two is minor, as demonstrated in Appendix~\ref{ap_vcut}.

Conceptually, this cut is implicitly making a distinction between
small-scale ``galactic fountain'' processes that occur at the disk-halo
interface (scales out to a few tens of $\mathrm{kpc}$ for a Milky Way-mass galaxy) over
timescales comparable to the galaxy dynamical time, and a 
larger-scale ``halo fountain'' processes (tens to hundreds of $\mathrm{kpc}$) that 
occur over timescales more comparable to a halo dynamical time (about
one tenth of a Hubble time). Small-scale galactic fountains
are poorly resolved in our analysis (due both to the finite time
resolution of our simulation outputs, and to the limited spatial
resolution of the simulations), and in any case act over timescales that
are too short to have a significant direct impact on the efficiency
with which galaxies form stars. As such our focus in this study is
on wind recycling associated with the larger-scale halo fountain (and
also on recycling of gas that moves outside the virial radius). Small-scale galactic fountains are
of interest in other contexts, for example as a fine-grained mechanism to bring
in mass and angular momentum from a hot corona \cite[e.g.][]{Fraternali17}.
Furthermore, observations of inflowing gas at the disk-halo interface \cite[e.g.][]{Bish19} may
be tracing smaller-scale galactic fountain processes that are either
explicitly removed, or are unresolved, in our analysis.

Particles that leave galaxies (or haloes), and that pass the time-integrated
velocity cut, are then tracked at later simulation snapshots, which enables
us to establish if accreted particles are being accreted onto galaxies (or 
onto haloes) as:
\begin{enumerate}
\item first-time accretion,
\item recycled accretion from a progenitor of the current galaxy (or halo),
\item transferred accretion that was previously inside the ISM (or halo) of a non-progenitor galaxy (or halo). 
\end{enumerate}
We also compute the mass of gas 
that has been ejected from progenitors of the present galaxy (or halo), and
that still currently resides outside the galaxy/halo. This is used to measure 
the characteristic efficiency of galaxy-scale and halo-scale wind recycling 
(see Section~\ref{rec_sec}).

We also separate inflow rates between ``smooth'' accretion (which can take
the form of any of the three afore-mentioned accretion modes) and mergers.
At the halo scale, particles that were inside the virial radius of a subhalo 
at the snapshot prior to being accreted onto the FoF group are considered 
as merging material. Note that this material can continue to be
associated with satellite subhaloes, at least while they can still be identified
in the simulation. Similarly, at the galaxy scale any gas that was
inside the ISM of a non-main progenitor galaxy at the prior snapshot
is considered as merging material. Any material that is not brought
in by mergers is instead considered ``smooth'' accretion. Note that for dark matter
(which we only measure at the halo scale), ``smooth'' accretion is
not an intrinsically well defined quantity, as essentially all
dark matter would, if simulated at infinite numerical resolution,
be accreted within haloes, depending precisely on the
cutoff scale of the matter power spectrum, which in turn depends 
on the nature of the dark matter particle. Modern cosmological 
simulations typically only have the mass resolution to resolve 
$\approx 50 \%$ of the total mass in haloes \cite[e.g.][]{Springel05,Genel10,VanDaalen15}. 
For gas, substantial true smooth
accretion would be expected independent of numerical resolution,
since after reionization the UVB (among other processes)
provides indirect pressure support that both removes and 
prevents gas from ever being accreted onto very low-mass 
haloes.

Here, we define ``smooth'' accretion by setting an explicit
fiducial halo mass cut at $9.7 \times 10^8 \, \mathrm{M_\odot}$,
corresponding to the mass of $100$ dark matter (numerical) 
particles at fiducial \eagle resolution.
Gas or dark matter that is accreted onto a host halo while within another
halo with mass lower than the limit are considered as smooth accretion.
In addition, we do not track particles that are ejected
from haloes (and their associated galaxies) below this mass scale, meaning that the limit
also affects our definitions of first-time, recycled,
and transferred (smooth) accretion. At the galaxy scale, we evaluate
the maximum past mass of satellite subhaloes, and only
count merging satellite galaxies to the merger accretion
rate if the maximum past subhalo mass exceeds the cut.
We assess the impact of varying our fiducial halo mass
cut for smooth accretion in Appendix~\ref{ap_smthr}, and
find that smooth gas accretion rates are generally well converged
at the chosen mass cut (but would not have been if we had used a higher
mass cut).

The methodology described here is similar in many respects to the
methods employed by studies of gas inflow and wind recycling
in cosmological zoom-in simulations 
\cite[e.g.][]{Ubler14,Christensen16,AnglesAlcazar17,Grand19}.
One noteworthy difference is in our definition of the
distinction between ``transferred'' and ``recycled'' accretion,
the potential importance of which has recently been highlighted
by \cite{AnglesAlcazar17} and \cite{Grand19}. These authors
implicitly consider ``transfer'' as consisting of
particles that were ejected from any galaxy that is 
not flagged as the main progenitor of the descendant
onto which the particles are now being accreted. Here, we instead define ``recycled''
accretion as particles that originated from any progenitor
of the current galaxy (or halo), meaning that ``transferred''
accretion must originate from a non-progenitor galaxy.
In practice, this means that gas that is ejected from
satellites and then reaccreted after the satellite
has merged with the host is tagged as recycled accretion
in our scheme, but would be considered as transferred
accretion in the afore-mentioned studies. While
ultimately subjective, we regard our
choice as being the more natural definition of wind
recycling (distinct from ``transfer''), since the definition 
of the main progenitor is often fairly arbitrary (though
admittedly by low redshift a single main progenitor
branch generally exists clearly within a merger tree), and
the product of a galaxy merger should be considered
as the sum of all progenitors. In addition, our definition 
of recycling naturally maps onto the framework of analytic and 
semi-analytic models, which generally merge the 
tracked ejected gas reservoirs of galaxies when
they merge. We show the impact of this choice
in Appendix~\ref{ap_mp} (our definition slightly
increases the importance of recycling
relative to transfer), and we take care to use consistent
definitions when comparing to other simulations
in Section~\ref{acc_mode_comp_sec}.

Finally, we do not attempt to establish the physical reason for
why ``transferred gas'' is removed from a galaxy before
being accreted onto another. Physical mechanisms may
include feedback-driven outflows, ram pressure stripping
against the corona of a massive host galaxy, or stripping by
gravitational tides. Disentangling these effects 
is a non-trivial problem that sits
outside the scope of our current study, but remains
an interesting avenue for future work \cite[see also][for work on stripping in \eagle]{Marasco16}.
Note that with our definition of ``transferred'' versus ``recycled'' accretion,
gas that is stripped from satellites will be labelled as recycled
accretion if the associated satellite merges before (or at the same time as) 
it is accreted onto the central galaxy, meaning that our definition of ``transferred''
gas may have a closer connection to feedback-driven outflows. 
We note also that when comparing our results to \cite{Grand19}, who
are able to cleanly separate stripped versus feedback-driven
transfer events due to their explicit wind model,
we group these two components together for the
purpose of the comparison.

\section{Results}
\label{results_section}

\begin{figure}
\includegraphics[width=20pc]{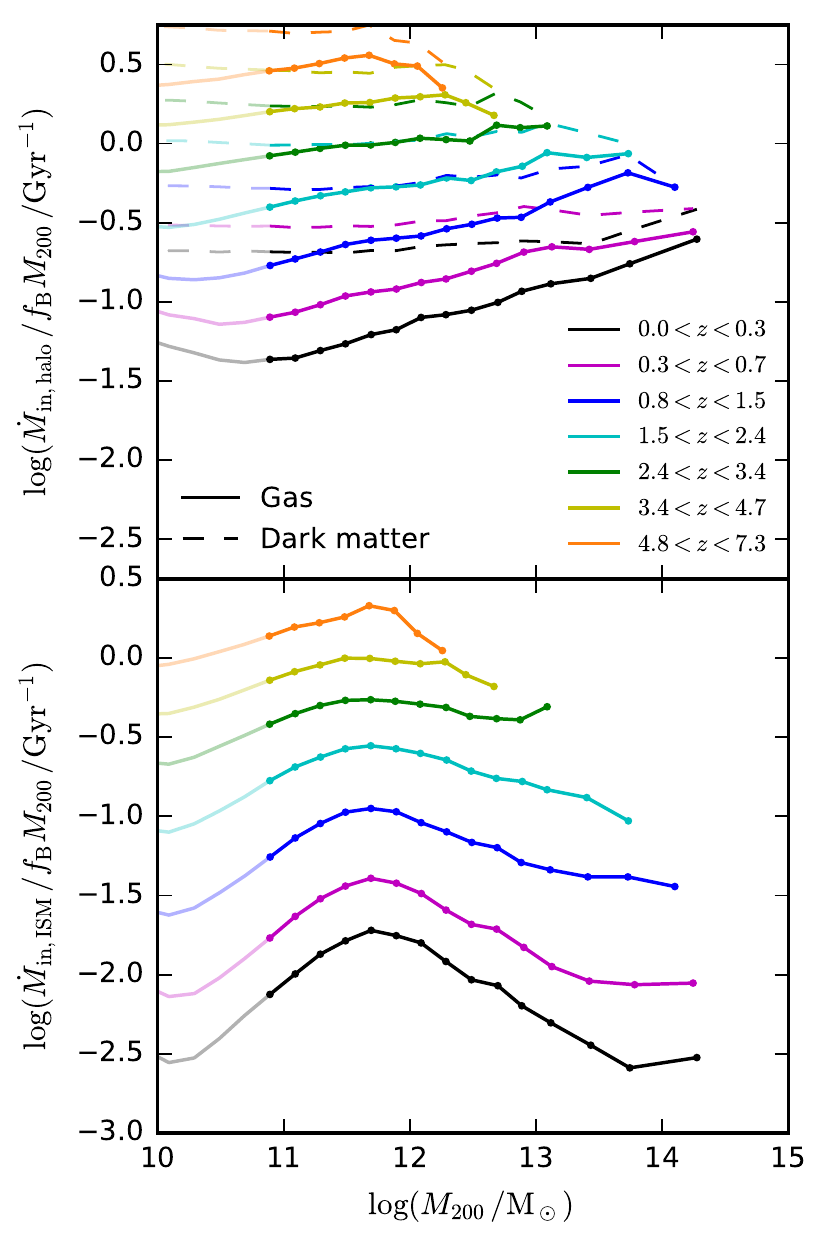}
\caption{Gaseous inflow rates through the halo virial radius (top), 
and onto the ISM (bottom) of central galaxies, as a function of halo mass.
Solid (dashed) lines show inflow rates for gas (dark matter).
Inflow rates are normalised by $f_{\mathrm{B}} M_{200}$ for gas, and by $(1-f_{\mathrm{B}}) M_{200}$ for dark matter, 
where $f_{\mathrm{B}} \equiv \frac{\Omega_{\mathrm{b}}}{\Omega_{\mathrm{m}}}$ is the cosmic baryon fraction,  
such that the normalised rates are equal if baryonic inflow perfectly traces the dark matter inflow rate at the virial radius.
Inflow rates include contributions from smooth accretion and halo/galaxy mergers.
Transparent lines indicate the range where there are fewer than $100$ stellar particles per galaxy.
}
\label{gas_accretion_figure}
\end{figure}

Fig.~\ref{gas_accretion_figure} shows the total inflow rates of gas and dark matter onto haloes (top panel), as
well as the total inflow rate of gas onto the ISM of galaxies (bottom panel), plotted as a function of halo mass. 
The values plotted here (and in all subsequent figures unless otherwise stated) are the average, 
which we compute \cite[following][]{Neistein12,Mitchell20} as the mean value of the numerator, 
divided by the mean value of denominator. This helps to ensure that the time-integral of the
average inflow rates sum to the correct value of the average of the individual time-integrated 
rates. We compute averages by combining simulation snapshots in the redshift intervals
indicated. Only central galaxies are included in the average 
\cite[see][for a detailed study of the differences between inflows onto centrals and satellites in \eagle]{Vandevoort17}.

Fig.~\ref{gas_accretion_figure} shows the expected basic behaviour for gaseous inflow
rates onto galaxies and haloes. At fixed mass, inflow rates increase with increasing
redshift, reflecting the overall decline in the average density of the Universe with
time. If we scale out the zeroth order time dependence by multiplying by the age
of the Universe (not shown, but see Fig.~\ref{infall_figure}), the redshift evolution is weaker, but there is
still approximately $0.5 \, \mathrm{dex}$ of evolution over $0<z<5$, with inflow rates still
increasing with increasing redshift. Since dark matter haloes approximately grow at a rate that scales  
inversely with the Hubble time (at fixed mass), this implies that there may be additional
processes beyond gravity alone that shape the redshift evolution of baryonic
accretion onto galaxies and haloes in the simulation. 
In detail however, dark matter growth rates are not expected to scale
exactly with the Hubble time \cite[e.g.][]{Correa15}, and indeed if 
we scale out the the age of the Universe, we do find that 
our measurements of dark matter accretion rates decrease by about $0.2 \, \mathrm{dex}$
over the same redshift interval for which gas accretion rates decline by 
$0.5 \, \mathrm{dex}$. This implies therefore that there may also be
effects related to pure gravitational evolution that affect the redshift evolution 
of gaseous inflow rates.

At fixed redshift, inflow rates increase with halo mass, both at the galaxy
scale and at the halo scale. We choose
to present results by first scaling out this zeroth order mass dependence,
both to compress the dynamic range and also to highlight the important change
in behaviour for galaxy-scale accretion at the characteristic halo mass
of $\sim 10^{12} \, \mathrm{M_\odot}$
\cite[see][to view inflow rates in \eagle without this rescaling]{Correa18b}.
Normalised galaxy-scale inflow rates (bottom panel) clearly peak at (slightly below)
the mass scale of $10^{12} \, \mathrm{M_\odot}$ in \eagle, though the feature becomes weaker with increasing
redshift. The feature has a clear and obvious connection to the shape of the
relationship between galaxy stellar mass and halo mass, in the
sense that the ratio of stellar mass to halo mass also peaks strongly
at the same characteristic halo mass \cite[e.g.][]{Behroozi10,Moster10}.
Interestingly, inflow rates at the halo scale do not show this peak 
(top panel), which aligns with the classic picture of galaxy formation
in which longer radiative cooling timescales in high-mass haloes
act to prevent coronal gas in the circum-galactic medium (CGM) from reaching 
the ISM, but not from being accreted onto the halo at the scale of the 
virial radius \cite[e.g.][]{Rees77}, although the
modern picture also requires effective AGN feedback to prevent a cooling
flow \cite[e.g.][]{Bower06,Croton06}, and these two ingredients may not be independent \cite[][]{Bower17}.
Fig.~\ref{gas_accretion_figure} shows that gaseous inflow rates at the virial radius do however fall short of
dark matter accretion rates (after scaling out the cosmic baryon
fraction) at lower halo masses. We return to this point in Section~\ref{prev_sec}.

\begin{figure*}
  \begin{center}
\includegraphics[width=40pc]{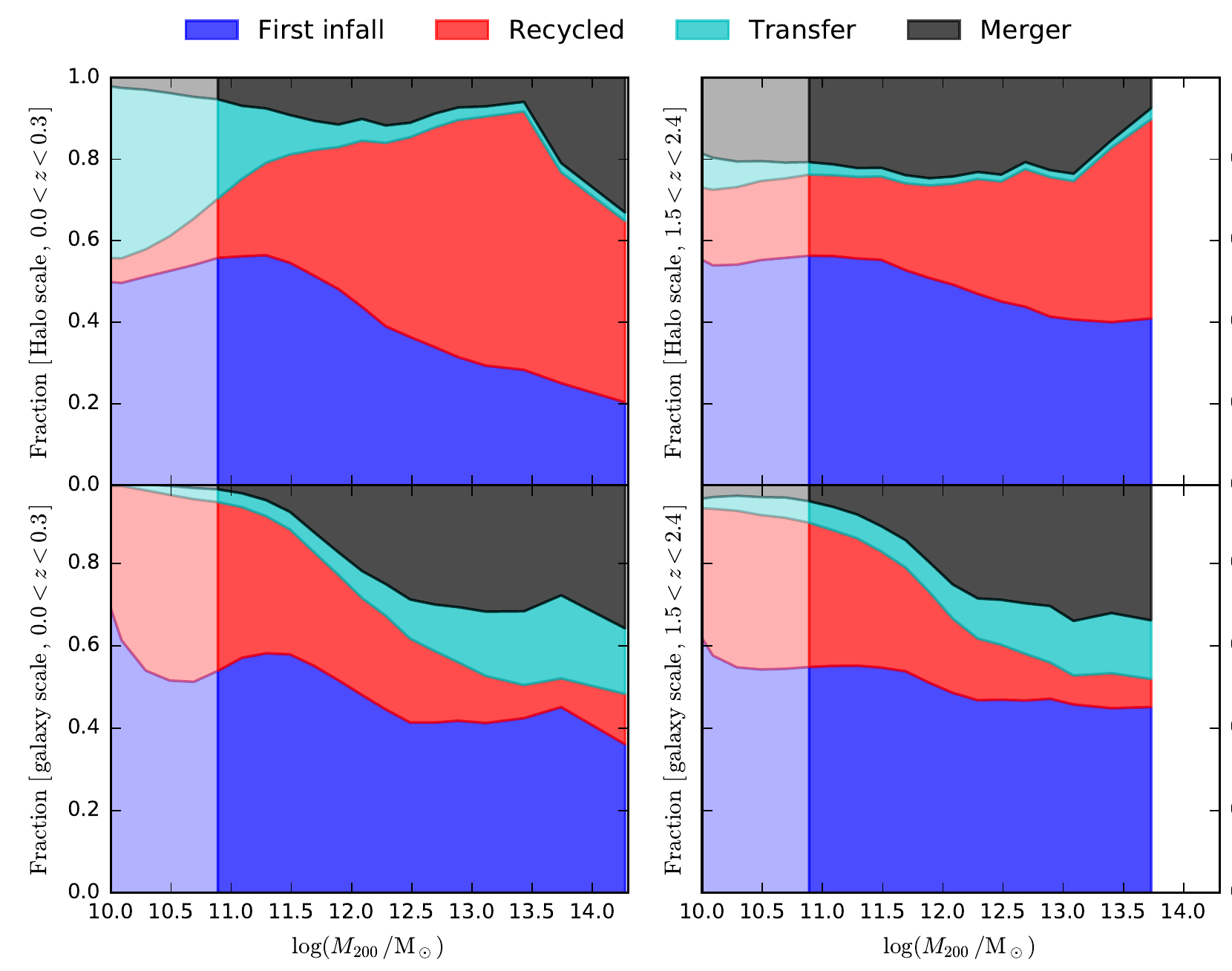}
\caption{The fractional contribution of different inflow components to the total inflow rate onto
  haloes (top panels), and onto the ISM of central galaxies (bottom panels), as a function of halo mass.
  The coloured shaded regions show the contributions from first-infall (blue), recycling (i.e. from progenitors, red), 
  transfer (i.e. from non-progenitors, cyan), and mergers (black).
  The definition of these components is distinct at the halo (top) and galaxy (bottom) scales,
  as described in the main text.
  Left (right) panels show averages for the redshift interval $0<z<0.3$ ($1.5 < z < 2.4$).
  The region with increased transparency indicates the halo mass range for which there are 
  fewer than $100$ stellar particles per galaxy.
}
\label{return_contr}
  \end{center}
\end{figure*}

Fig~\ref{return_contr} shows the relative contribution of smooth gaseous accretion, split
between first-time infall, recycled infall (from progenitors of the current
galaxy/halo), and transferred gas (from non-progenitors of the current galaxy/halo), 
as well as the contribution from mergers (at the halo scale this refers
to the accretion of satellite subhaloes through the virial radius of the host). 
At both galaxy and halo scales, the single most important contributor to total 
gas accretion is generally provided by gas that is infalling for the first time.
Mergers become an important source of gaseous accretion in higher-mass haloes
($M_{200} \geq 10^{12} \, \mathrm{M_\odot}$), especially at the galaxy scale (bottom panels).
At the halo scale (top panels), recycling plays an important role at high halo masses, and
actually provides the largest individual contribution for $M_{200} > 10^{12} \, \mathrm{M_\odot}$ in the
low-redshift interval plotted. 
This trend is reversed at the galaxy scale however, with galaxy-scale recycling playing 
the largest role in lower-mass haloes 
($M_{200} \approx 10^{11} \, \mathrm{M_\odot}$), and is subdominant in group 
and cluster mass haloes. Transferred accretion is negligible at the halo
scale, aside from for very low-mass haloes ($M_{200} < 10^{11} \, \mathrm{M_\odot}$).
Transferred accretion does play a role in higher mass haloes at the galaxy scale 
however, providing up to $20 \%$ of the total gas accretion for $M_{200} > 10^{12} \, \mathrm{M_\odot}$.

While not shown, we have computed the mass fraction of stars that form in galaxies
from the different accretion channels discussed here. In principle, stars
may form from a biased sub-sample of the accreted gas; for example one
could envisage that recycled gas is more metal enriched, and so more
readily able to form stars. We find however that star formation associated
with the different accretion channels closely tracks inflow rates 
onto galaxies, with no obvious bias favouring a particular accretion channel.

Putting this together, we find that all of the accretion channels considered
play an important role for at least a subset of the various mass and spatial scales
considered. Furthermore, we find that different individual accretion components scale in qualitatively
distinct ways with with halo mass (not shown for conciseness).
For example, the peak in the total (halo mass normalised) galaxy-scale gaseous accretion rates
seen in Fig.~\ref{gas_accretion_figure} at $M_{200} \sim 10^{12} \, \mathrm{M_\odot}$
is primarily created by the recycled and first-infalling components, and is
not associated with the transfer and merging components. 
Overall, the situation is complex, reflecting the physics
of radiative cooling and feedback on different scales. We now proceed
to focus on isolating different parts of this picture in the following 
parts of this section.

\subsection{Preventative feedback}
\label{prev_sec}

The top panel of Fig.~\ref{gas_accretion_figure} shows that 
inflow rates at the virial radius fall short of
dark matter accretion rates (after scaling out the cosmic baryon
fraction) in low-mass haloes, with the discrepancy between
the two growing with decreasing halo mass. At low redshift, 
the discrepancy actually becomes smaller at very low halo masses 
($M_{200} < 10^{11} \, \mathrm{M_\odot}$), but this is the 
regime where galaxies are poorly resolved in \eagle. 
This feature aside, the decline in gaseous inflow compared 
to dark matter inflow at lower halo masses can be 
attributed to the impact of feedback
processes, as demonstrated explicitly in the recent
dedicated study of halo-scale accretion in \eagle by 
\cite{Wright20}. This ``preventative feedback'' effect (at
the halo scale) has also been noted in studies of other 
cosmological simulations \cite[][]{FaucherGiguere11,VanDeVoort11,Christensen16, Mitchell18_ramses, Tollet19}.
Generally speaking, we find that \eagle predicts larger offsets between
gaseous and dark matter accretion rates than
in other simulations, most
notably for haloes in the mass range 
$10^{12} < M_{200} / \, \mathrm{M_\odot} < 10^{13}$
\cite[where there is no effect of feedback on halo-scale accretion
in the OWLS simulations, for instance,][]{VanDeVoort11}.

An important question, which has not (to our knowledge)
been addressed in previous studies, is whether preventative
feedback at the halo scale (if quantified
as the ratio of the rates of total gas accretion to total dark matter
accretion) reflects a reduction in smooth gas accretion
onto haloes (either because of the ram pressure of outflows,
or results from the thermal pressure injected into the CGM
and intergalactic medium by feedback), or instead simply reflects the
removal of baryons from progenitor haloes before they are accreted
onto the main progenitor branch of the descendant halo
\cite[similar to the concept of pre-processing of satellite
galaxies in group-mass haloes before falling into
galaxy clusters,][]{Bahe13}.

In \cite{Mitchell20}, we show that feedback drives large-scale 
outflows at the scale of the virial radius in \eagle,
which will indeed therefore reduce the baryon content of accreted
satellite subhaloes before they are accreted through the virial radius of the host.
Here, we focus instead on the question of preventative
feedback acting via the reduction of smooth gaseous accretion
(defining smooth accretion as any gas or dark matter
that enters the halo while \textbf{not} within the virial radius
of a smaller halo of mass $M_{200} > 9.7 \times 10^8 \, \mathrm{M_\odot}$, corresponding
to $100$ dark matter particles at standard \eagle resolution).

\begin{figure}
\includegraphics[width=20pc]{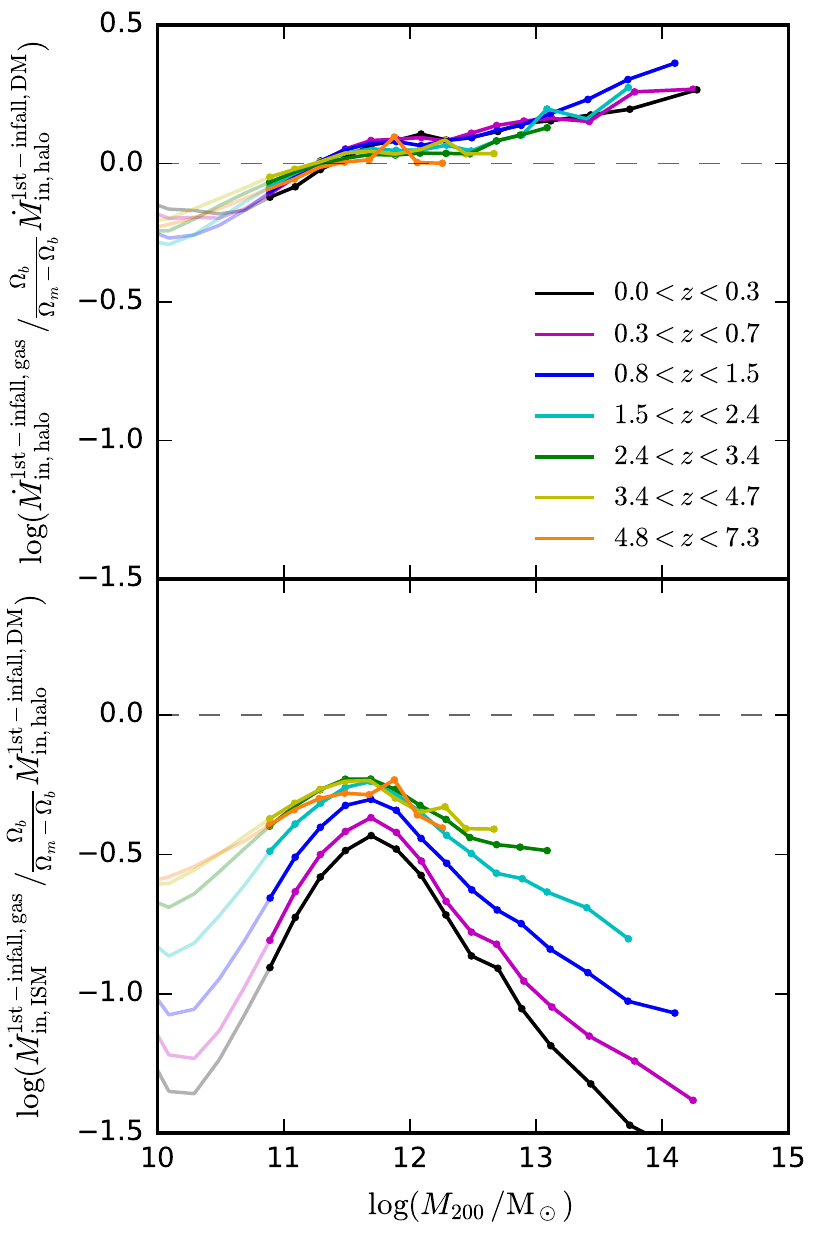}
\caption{The ratio of the rates of smooth first-time gas inflow to smooth first-time dark matter inflow, as a function of halo mass.
The top panel shows the ratio for gas to (scaled) dark matter inflow measured at the halo virial radius.
The bottom panel shows the ratio of gas accretion rate onto the ISM of central galaxies, 
divided by the scaled dark matter inflow rate at the halo virial radius.
Transparent lines indicate the range where there are fewer than $100$ stellar particles per galaxy.
At the virial radius, first-time gas accretion is slightly suppressed relative to first-time dark matter 
accretion at the virial radius for $M_{200} < 10^{11} \, \mathrm{M_\odot}$, but is actually 
enhanced relative to dark matter accretion at higher masses (note however that
total gas accretion rates are always comparable or lower than total dark matter accretion rates,
see \protect Fig.~\ref{gas_accretion_figure}).
First-time gas accretion onto the ISM of galaxies is always suppressed relative to first-time
dark matter accretion at the virial radius, and more so at low-redshift, and at both low and high
halo masses.
}
\label{prev}
\end{figure}

The standard assumption for gaseous accretion (as exemplified
for example by semi-analytic galaxy formation models) is that
smoothly accreted gas that is being accreted onto a halo for
the first time will trace the equivalent for dark matter.
We therefore plot the ratio of first-time infall rates of gas and dark
matter onto haloes in the top panel of Fig.~\ref{prev}, rescaling the dark matter
rate by $\Omega_b / (\Omega_m - \Omega_b)$.
The grey horizontal dashed line therefore indicates the expected value if gas traces dark matter.
For low-mass haloes with $M_{200} < 10^{11} \, \mathrm{M_\odot}$,
we see that the ratio is below unity, implying that smooth 
gas accretion is indeed reduced compared to dark matter accretion.
Intriguingly, the opposite is true for $M_{200} > 10^{12} \, \mathrm{M_\odot}$,
for which first-time infalling gas is being more efficiently
accreted onto haloes, by up to $0.3 \, \mathrm{dex}$.

\begin{figure}
\includegraphics[width=20pc]{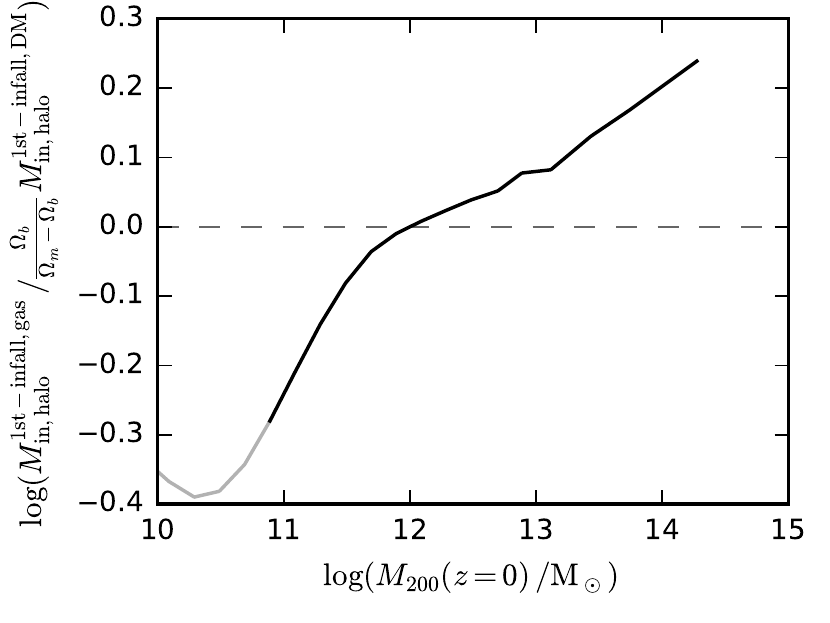}
\caption{The ratio of masses of time-integrated first-time accreted gas over time-integrated first-time accreted dark matter, 
  for matter accreted onto haloes by $z=0$, as a function of the final halo mass. 
  First-time accretion is computed across all progenitors of the
  final halo. Dark matter accretion is scaled by $\frac{\Omega_b}{\Omega_m-\Omega_b}$, 
  and the grey horizontal dashed line indicates the expected value if gas traces dark matter.
  Transparent lines indicate the range where there are fewer than $100$ stellar particles per galaxy.
  Integrated over the entire history of a halo, first-time gas accretion closely traces dark matter for
  $M_{200} \sim 10^{12} \, \mathrm{M_\odot}$, 
  is suppressed relative to dark matter accretion by up to $0.4 \, \mathrm{dex}$ at lower halo masses, 
  and exceeds dark matter accretion by $0.2 \, \mathrm{dex}$ for $M_{200} \sim 10^{14} \, \mathrm{M_\odot}$.
  The effect of preventative feedback is therefore stronger when integrated over the history
    of a halo, compared to the instantaneous view presented in \protect Fig.~\ref{prev}.
}
\label{prev_z0}
\end{figure}

An alternative perspective on this is presented Fig.~\ref{prev_z0}, which
compares the cumulative masses of first-time infalling gas and dark matter, 
integrated in time over the entire merger tree of each descendant halo. 
Viewed in this way, there is still an enhancement of
first-time gas accretion relative to dark matter in high-mass haloes ($M_{200} \gtrsim 10^{13} \, \mathrm{M_\odot}$), 
but the effect is weaker than seen in the instantaneous measure of preventative 
feedback presented in Fig.~\ref{prev}. Conversely, first-time gas accretion
is more suppressed relative to dark matter accretion for low-mass haloes in the integrated
measurement than for the instantaneous measurement 
(up to $0.4 \, \mathrm{dex}$ for $M_{200} \sim 10^{10} \, \mathrm{M_\odot}$).

\begin{figure*}
  \begin{center}
\includegraphics[width=40pc]{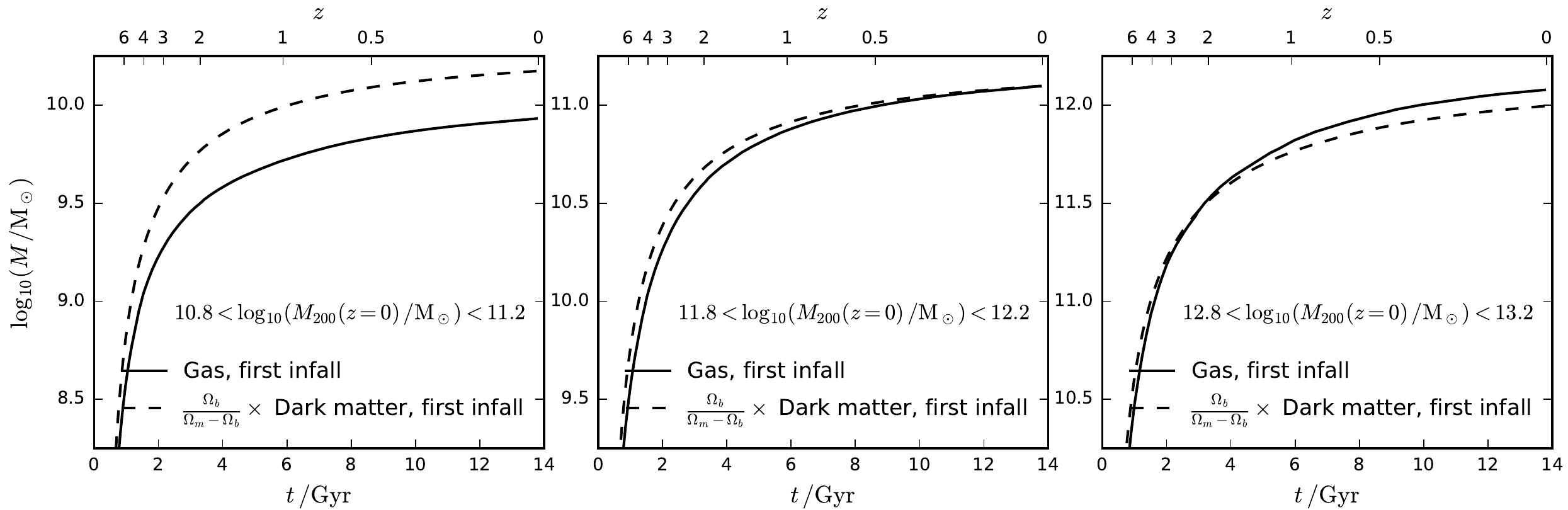}
\caption{The evolution of cumulative mass in first-time accreted gas (solid) and dark matter (dashed) 
  onto haloes, binned by the final halo mass at $z=0$, as labelled.
  First-time accretion is computed across all progenitors of the final halo.
  Dark matter accretion is scaled by $\frac{\Omega_b}{\Omega_m-\Omega_b}$.
  For the $M_{200}(z=0) = 10^{11} \, \mathrm{M_\odot}$ bin (left panel), gas accretion
  is consistently suppressed relative to dark matter for all but the earliest times.
  For the higher halo mass bins, gas accretion rates are suppressed
  relative to dark matter at high redshift, but are enhanced relative to dark matter
  at lower redshifts. For the $M_{200}(z=0) = 10^{12} \, \mathrm{M_\odot}$ bin (middle panel), 
  this has the outcome that the cumulative first-time gaseous and dark matter accretion
  balance by $z=0$. Feedback therefore has the effect of (slightly) delaying gas accretion onto 
  haloes for this mass range, without preventing any of the total expected amount of gas 
  accretion by $z=0$.
}
\label{prev_evo}
  \end{center}
\end{figure*}

Fig.~\ref{prev_evo} then completes the picture by presenting the time evolution
of the cumulative mass accretion of first-infalling gas and dark matter,
again integrating over all progenitors of descendant haloes that are binned
in mass at $z=0$. Smooth first-time gas accretion
is always suppressed relative to dark matter for the progenitors of haloes
of mass $M_{200}(z=0) = 10^{11} \, \mathrm{M_\odot}$. 
At $M_{200}(z=0) = 10^{12} \, \mathrm{M_\odot}$, gas
is delayed from being accreted onto the halo, rather than being prevented
from entering (by $z=0$). This has the effect of shifting the peak redshift 
for instantaneous halo gas accretion rates (not shown) from $z \approx 4$ 
(as for the dark matter) to $z \approx 3$.
At $M_{200}(z=0) = 10^{13} \, \mathrm{M_\odot}$, gas accretion is reduced 
at high redshift, but the cumulative mass
is slightly enhanced compared to that of dark matter for $z<2$. We
speculate that this could result from the enhanced radiative
cooling rates that are possible once feedback enriches gas
in the halo outskirts with heavy elements.

Preventative feedback has also been explored 
at the galaxy scale \cite[e.g.][]{FaucherGiguere11,VanDeVoort11,Dave12},
at which point the term refers to the combined effects of feedback slowing or
stopping the rates of gaseous infall on circum-galactic
(and larger) scales, as well as the long-predicted
effect that gas infall onto galaxies is restricted
by long radiative cooling timescales in the coronae of
high-mass haloes.
The bottom panel of Fig.~\ref{prev} presents an
instantaneous measure of preventative feedback when
framed in this way, plotting the ratio of first-time
gaseous infall onto the ISM of galaxies, relative to the
first-time infall of dark matter onto dark matter haloes.
The results echo the trends seen in
Fig.~\ref{gas_accretion_figure}, showing that gas
accretion onto galaxies peaks strongly at 
$M_{200} \sim 10^{12} \, \mathrm{M_\odot}$.
This partly reflects the afore-mentioned preventative
feedback at the virial radius, but also reflects
the efficiency with which gas is able to infall from
the CGM (within the virial radius) down onto the ISM.
We separate the latter effect in the next section.

\subsection{Infall from the CGM}
\label{infall_subsec}

\begin{figure}
\includegraphics[width=20pc]{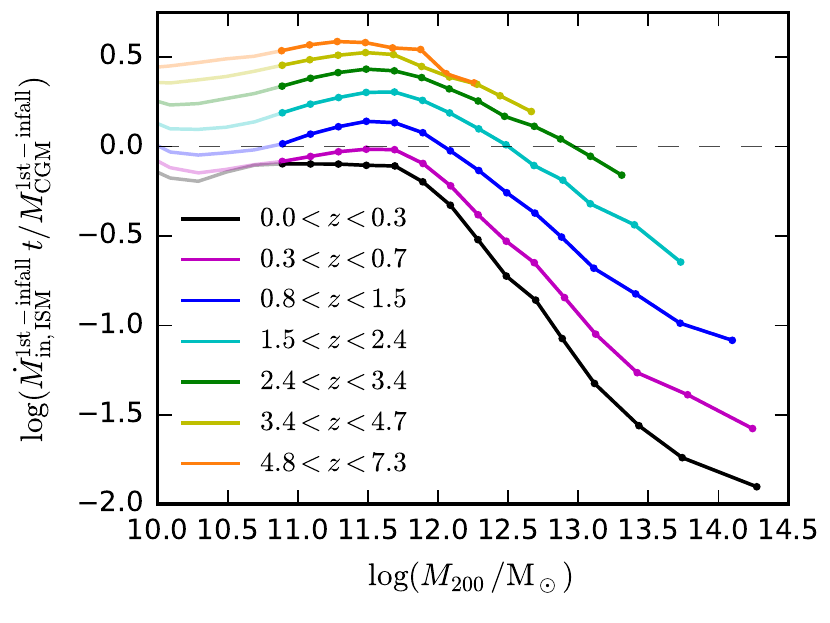}
\caption{Accretion rates of gas inflowing for the first time onto the ISM of central galaxies, 
  normalised by the mass in the CGM out to the virial radius, as a function of halo mass.
  For both quantities, only gas that has never been in the ISM of a galaxy is included.
  We also scale out the age of the Universe, $t$, at each redshift.
  With the chosen normalisation, the normalised inflow rates define a dimensionless efficiency 
  for gas to infall from the CGM onto the ISM for the first time
  (a value of unity implies that gas infalls over a Hubble time).
  Transparent lines indicate the range where there are fewer than $100$ stellar particles per galaxy.
  The infall efficiency increases weakly with halo mass for $M_{200} < 10^{12} \, \mathrm{M_\odot}$, 
  peaking slightly at this mass, 
  and then strongly decreases with halo mass for $M_{200} > 10^{12} \, \mathrm{M_\odot}$.
  This reflects the transition from short to long radiative cooling times due to the shift
  of the virial temperature beyond the peak of the cooling curve, and the effect of AGN feedback.}
\label{infall_figure}
\end{figure}

Fig.~\ref{infall_figure} presents the efficiency of first-time gaseous infall from the CGM onto the ISM.
We define this efficiency as 
$\dot{M}_{\mathrm{in,ISM}}^{\mathrm{1st-infall}} / M_{\mathrm{CGM}}^{\mathrm{1st-infall}}$,
where $\dot{M}_{\mathrm{in,ISM}}^{\mathrm{1st-infall}}$ is the inflow rate of gas onto the ISM for
gas that has never been in the ISM of a galaxy before, and $M_{\mathrm{CGM}}^{\mathrm{1st-infall}}$
is the mass of gas in the CGM (by which we mean outside the ISM but within the halo virial radius) of
the central subhalo, and that also has never been in the ISM of a galaxy before. This efficiency
is the inverse of the characteristic timescale for the first-infalling gas in the CGM to be
depleted onto the ISM (exactly analogous to the standard definition of the gas depletion time
in the ISM, for example). We then scale out the zeroth order time dependence by multiplying
by the age of the Universe, which defines a dimensionless efficiency of first-time CGM infall.

Fig.~\ref{infall_figure} shows that the efficiency of first-time infall from the CGM is nearly
(but not completely) independent of halo mass for $M_{200} < 10^{12} \, \mathrm{M_\odot}$, but
declines strongly with increasing halo mass for $M_{200} > 10^{12} \, \mathrm{M_\odot}$.
This again reflects the classic anticipated dichotomy in galaxy formation between a regime
in which infall is limited primarily by gravitational timescales (which are scale free,
and so independent of halo mass) in low-mass haloes, to a regime where infall is limited
by radiative cooling timescales (which are strongly scale dependent, due to atomic physics)
and AGN feedback.
Intriguingly, the efficiency of first-time infall does peak slightly at 
$\sim 10^{12} \, \mathrm{M_\odot}$ at higher redshifts (but not at $z=0$), indicating
that there may be more than just gravity regulating infall, even in the limit of short
cooling times. Galaxy star formation rates, and therefore outflow rates, also increase strongly with
increasing redshift \cite[e.g.][]{Mitchell20}, and so we speculate that the slight
decrease in the first-time infall efficiency with decreasing halo mass for 
$M_{200} < 10^{12} \, \mathrm{M_\odot}$
for $z>1$ could be related to feedback processes. This is however a smaller effect compared to the
modulation of first-time infall on larger scales discussed in Section~\ref{prev_sec},
implying that feedback may primarily regulate first-time gas infall on larger spatial scales. 

This can be rationalised by supposing that gas inflow rates in low-mass haloes are 
comparatively less affected by feedback 
in the inner CGM, as outflows propagate perpendicular to
the disk \cite[which we have shown to the case in \eagle, ][]{Mitchell20}, away from 
the primary plane of small-scale inflow, while outflows and inflows are more
isotropic (and so can more readily interact) on larger scales beyond the virial radius.
Note that galactic winds do evacuate significant amounts of gas out of the halo
virial radius (most of which has never been in the ISM before), affecting the 
denominator of our defined infall efficiency. As seen in the lower-right panel of
figure 1 in \cite{Mitchell20}, the (halo mass-normalised) outflow rate at the virial
radius is approximately independent of halo mass, meaning that this effect
will not affect the halo-mass dependence of the infall efficiency shown here.

Considering instead the redshift evolution of the first-time infall efficiency,
Fig.~\ref{infall_figure} shows that even after scaling
out the zeroth order time dependence, there is still about $0.5 \, \mathrm{dex}$
of evolution at fixed mass over the interval $0<z<3$. This is contrary
to the expectation that (in the regime of short radiative cooling times)
gas infalls from the CGM over a gravitational freefall timescale, since
this timescale scales approximately with the halo dynamical time
($t_{\mathrm{dyn}} \equiv R_{\mathrm{vir}} / V_{\mathrm{circ}}$, where
$V_{\mathrm{circ}}$ is the halo circular velocity at $R_{\mathrm{vir}}$), 
which itself is approximately $10 \, \%$ of the Hubble time at a given redshift.
The decrease in the infall efficiency with decreasing redshift could reflect
the increase in specific angular momentum with decreasing redshift at
fixed halo mass (providing more rotational support against collapse to the center),
or could be related to the impact of feedback processes, either by reducing
inflow rates directly (numerator of the infall efficiency), or by altering 
the overall mass reservoir of the CGM (denominator of the infall efficiency).
At higher halo masses, where radiative cooling is expected to provide
the limiting timescale, lower infall efficiencies can be straightforwardly
explained by the lower average densities at low-redshift \cite[e.g.][]{Correa18b}.

\subsection{Galactic and halo-scale wind recycling}
\label{rec_sec}

\begin{figure}
\includegraphics[width=20pc]{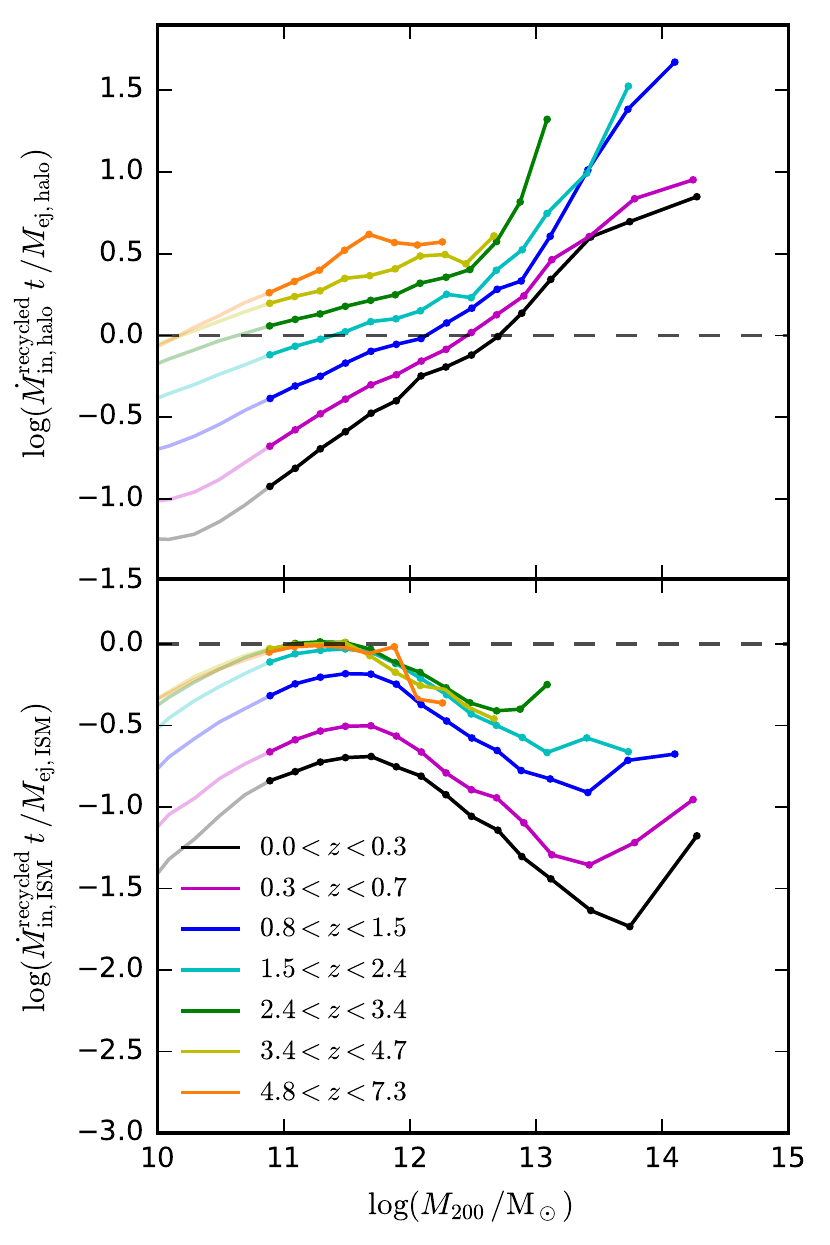}
\caption{Inflow rates of recycled gas through the halo virial radius (top panel),
  and onto the ISM of central galaxies (bottom panel), as a function of halo mass.
  The inflow rates in the top (bottom) panel are normalised by the mass of material 
  that has been ejected from progenitor haloes (galaxies), and that
  still currently resides outside the virial radius (ISM) at the plotted redshift.
  Inflow rates are multiplied by the age of the Universe at each redshift,
  altogether defining a dimensionless efficiency of wind recycling at the
  halo or galaxy scale.
  Transparent lines indicate the range where there are fewer than $100$ stellar particles per galaxy.
  At $z=0$, halo-scale gas recycling is relatively efficient (timescales shorter than a Hubble time) 
  for $M_{\mathrm{200}} > 10^{13} \, \mathrm{M_\odot}$ at $z=0$, 
  but is inefficient for lower-mass haloes. Halo gas recycling becomes more efficient at 
  higher redshifts. Gas recycling onto galaxies is inefficient (timescales equal or longer than a Hubble time)
  for all masses/redshifts, and the efficiency peaks at $M_{200} \sim 10^{11.7} \, \mathrm{M_\odot}$.
}
\label{rec_eff_fig}
\end{figure}

We parametrise the efficiency of recycling for gas that is ejected from galaxies and haloes 
by measuring $\dot{M}_{\mathrm{in}}^{\mathrm{recycled}} / M_{\mathrm{ej}}$, where 
$\dot{M}_{\mathrm{in}}^{\mathrm{recycled}}$ is the rate of return of recycled gas, and $M_{\mathrm{ej}}$
is the instantaneous mass of the reservoir of ejected gas (which is currently
located outside the galaxy or halo). This definition gives the inverse of
the depletion time for the ejected gas reservoir. We then scale out the zeroth
order time dependence by multiplying by the age of the Universe, yielding
a dimensionless efficiency. We provide measurements at both galaxy and halo scales.
The galaxy-scale measurement includes gas that has been ejected from the ISM of galaxies
(irrespective of whether that gas is also ejected through the virial radius).
The halo-scale measurement includes gas that has been ejected beyond the halo virial radius
(in this case irrespective of whether that gas has ever been situated inside the 
ISM of a galaxy in the past). Note that the halo-scale measurement
is equivalent to the definition that is generally used in semi-analytic galaxy formation models;
we compare our measurements with the values adopted in such models
in Section~\ref{sam_comp_sec}.

The measurements of wind recycling efficiency are presented in Fig.~\ref{rec_eff_fig}. At the halo
scale (top panel), the recycling efficiency always increases with 
halo mass, approximately as
$\dot{M}_{\mathrm{in}}^{\mathrm{recycled}} / M_{\mathrm{ej}} \propto M_{200}^{0.6}$
at $z=0$. At the galaxy scale (bottom panel), the recycling efficiency
peaks at the characteristic mass scale of $10^{12} \, \mathrm{M_\odot}$.
At a fixed halo mass of $M_{200} = 10^{12} \, \mathrm{M_\odot}$, the 
dimensionless efficiency of recycling (at both galaxy and halo scales) decreases by nearly one order 
of magnitude from $z=3$ to $z=0$. At higher redshifts, the halo-scale
efficiency continues to increase, but there is no longer any clear evolution at
the galaxy scale.

Comparing the two recycling efficiencies, recycling is much more efficient at the halo
scale than at the galaxy scale for $M_{200} > 10^{12} \, \mathrm{M_\odot}$, but
is more comparable at lower halo masses. Compared to the efficiency corresponding
to characteristic gas return over a Hubble time (a value of unity, shown by
the dashed horizontal lines), the gas ejected at the halo scale typically
returns after a halo dynamical time (about one tenth of the age of
the Universe) for $M_{200} \approx 10^{13.5} \, \mathrm{M_\odot}$, and returns after a Hubble
time for $M_{200} \approx 10^{12} \, \mathrm{M_\odot}$ at $z=1$.
At the galaxy scale, ejected gas on average returns over a timescale
that is equal or longer than the Hubble time for all halo masses, reaching up to
ten times the Hubble time at $M_{200} \approx 10^{10} \, \mathrm{M_\odot}$
and at $M_{200} \approx 10^{13} \, \mathrm{M_\odot}$. Note however
that despite the very low efficiency of galaxy-scale wind recycling,
this still forms an important contribution to galaxy-scale inflow
rates, especially for $M_{200} \sim 10^{11} \, \mathrm{M_\odot}$
(Fig.~\ref{return_contr}).
This reflects the global inefficiency of gaseous inflow onto galaxies
in the simulation.

\begin{figure}
\includegraphics[width=20pc]{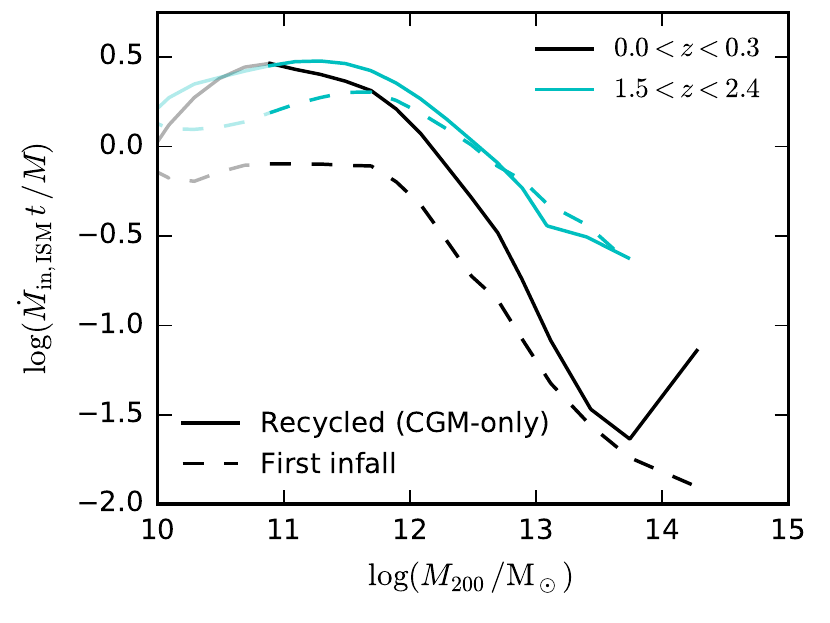}
\caption{The dimensionless efficiency of galaxy-scale wind recycling (solid
  lines), in this case measured relative to the subset of ejected gas that is still located within 
  the halo virial radius, with mass $M_{\mathrm{ej}}(r<R_{\mathrm{200}})$.
  We define the efficiency in this case as 
  $\dot{M}_{\mathrm{in,ISM}}^{\mathrm{recycled}} \, t \, / M_{\mathrm{ej}}(r<R_{\mathrm{200}})$,
  where $\dot{M}_{\mathrm{in}}^{\mathrm{recycled}}$ is the inflow rate of
  recycled gas onto the ISM, and $t$ is the age of the Universe.
  As a reference, we overplot the efficiency of first-time infall
  from the CGM onto the ISM (dashed lines), defined 
  as $\dot{M}_{\mathrm{in,ISM}}^{\mathrm{1st-infall}} \, t \, / M_{\mathrm{CGM}}^{\mathrm{1st-infall}}$,
  as introduced in \protect Fig.~\ref{infall_figure}.
  While global galaxy-scale gas recycling is always inefficient 
  in \eagle (Fig.~\ref{rec_eff_fig}), ejected gas that is still within $R_{200}$
  is recycled more efficiently, with an efficiency that is comparable or greater
  than that of first-time infall.
}
\label{rec_eff_fig_cgm}
\end{figure}

Importantly, much of the gas that is ejected from the galaxies in
\eagle is also ejected beyond the halo virial radius. As such, the
low efficiency we find for galaxy-scale wind recycling does not
necessarily imply that recycling is inefficient for the subset
of gas that is retained inside of the virial radius. This is
demonstrated in Fig.~\ref{rec_eff_fig_cgm}, which shows an alternative measure 
of the dimensionless efficiency of galaxy-scale wind recycling, in this case
defined as $\dot{M}_{\mathrm{in,ISM}}^{\mathrm{recycled}} \, t \, / M_{\mathrm{ej}}(r<R_{\mathrm{200}})$,
where $M_{\mathrm{ej}}(r<R_{\mathrm{200}})$ is now the mass of
ejected gas that is still within $R_{200}$. Defined in this
way, recycling of circum-galatic gas onto the ISM (solid lines)
is actually comparable or greater (by up to $0.5 \, \mathrm{dex}$)
in efficiency than that of first-time infall from the CGM 
(as introduced in Section~\ref{infall_subsec}, and shown as dashed lines here for comparison).
This illustrates that a subset of the gas that is ejected from
the ISM in \eagle is being efficiently (at least comparatively) recycled in a fountain
flow.

As an aside, we note that semi-analytic galaxy formation models
often implicitly assume that the efficiency of infall from the CGM (within
the virial radius) onto the ISM is the same for recycled and
first-infalling gas. Specifically, these models typically assume
that ``reincorporated'' ejected gas should be placed back in
the total reservoir of gas within $R_{200}$, which can then infall
onto the ISM with a single efficiency. Fig.~\ref{rec_eff_fig_cgm} 
shows that this is not an unreasonable assumption, since
the efficiency of CGM-scale recycling and first-infall
are qualitatively similar. Quantitatively, it may be
worthwhile for galaxy formation models to account for the possibility
that recycled gas is able to infall back onto the ISM with
a higher efficiency that that of gas that is infalling for
the first time, particularly by low-redshift.
Note that as discussed in Section~\ref{tracking_sec}, our definition
of galactic outflows (which are required to move outwards
a given distance over a given interval) means that this relatively 
high recycling efficiency is associated with a fountain
that, for the case of a Milky Way-mass halo at low redshift,
generally extends over scales of many tens of $\mathrm{kpc}$,
as opposed to a small-scale galactic fountain that operates
on scales of $10 \, \mathrm{kpc}$ or less.

\begin{figure}
\includegraphics[width=20pc]{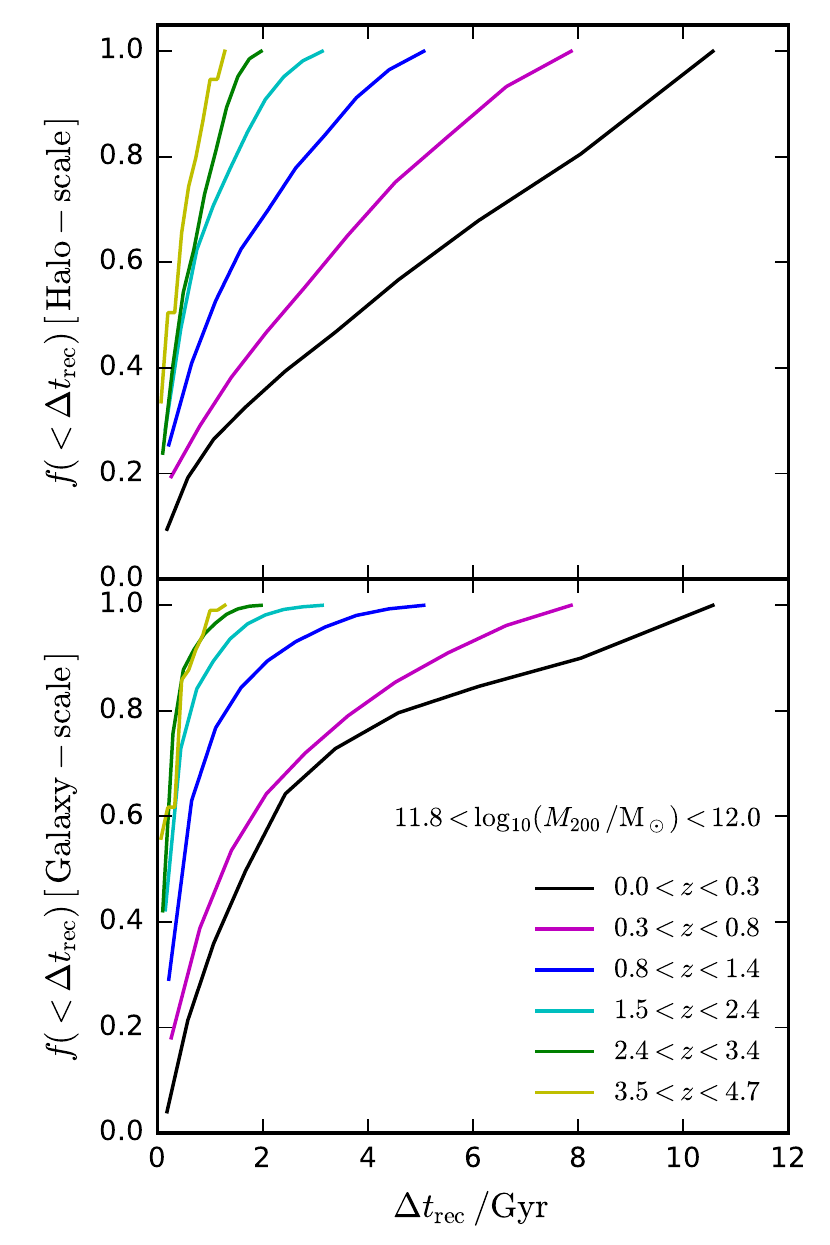}
\caption{The cumulative distribution of gas return timescales for gas being recycled at the indicated redshifts, 
  having been ejected at a time $\Delta t_{\mathrm{recycle}}$ in the past.
  The top (bottom) panel shows recycling onto haloes (the ISM of central galaxies), 
  selected with $10^{11.8} < M_{200} \, / \mathrm{M_\odot} < 10^{12}$.
  The distributions are plotted as the mean for the redshift bins indicated.
  The median return timescale of returning gas decreases with redshift, and
  is always shorter (and with a narrower distribution) at the galaxy scale than at the halo scale.
}
\label{tret_fig}
\end{figure}

Fig.~\ref{tret_fig} presents the cumulative residency time (time since ejection) 
distributions for ejected gas that is currently (i.e. at the redshift indicated) being recycled onto haloes (top panel), 
or onto the ISM of central galaxies (bottom panel). As before, the halo-scale residency 
time distribution includes gas irrespective of whether it was ejected from the ISM 
of a galaxy in the past. 
Distributions are plotted for a fixed halo mass range of $10^{11.8} < M_{200} \, /\mathrm{M_\odot} < 10^{12}$.

At the galaxy scale, the median recycling time at $0<z<0.3$ is $1.7 \, \mathrm{Gyr}$. Note that
this only accounts for returning gas, unlike the definition of the characteristic
return time plotted in Figs.~\ref{rec_eff_fig} and \ref{rec_eff_fig_cgm}.
The median recycling time depends strongly on redshift, reducing to
$480 \, \mathrm{Myr}$ by $z=1$, and to $230 \, \mathrm{Myr}$ by $z=2$.
This evolution reflects the evolution of the characteristic
halo dynamical time (which in turn shapes characteristic gravitational
freefall timescale), but may also reflect the development of a hot, pressurized
gaseous halo by low redshift, which could plausibly lengthen recycling
times.
Median recycling times are longer at the halo scale ($3.9 \, \mathrm{Gyr}$ at $0<z<0.3$),
and the distributions are broader, with a larger fraction of gas
returning after having been ejected from the halo at high redshift.

\begin{figure*}
\begin{center}
\includegraphics[width=40pc]{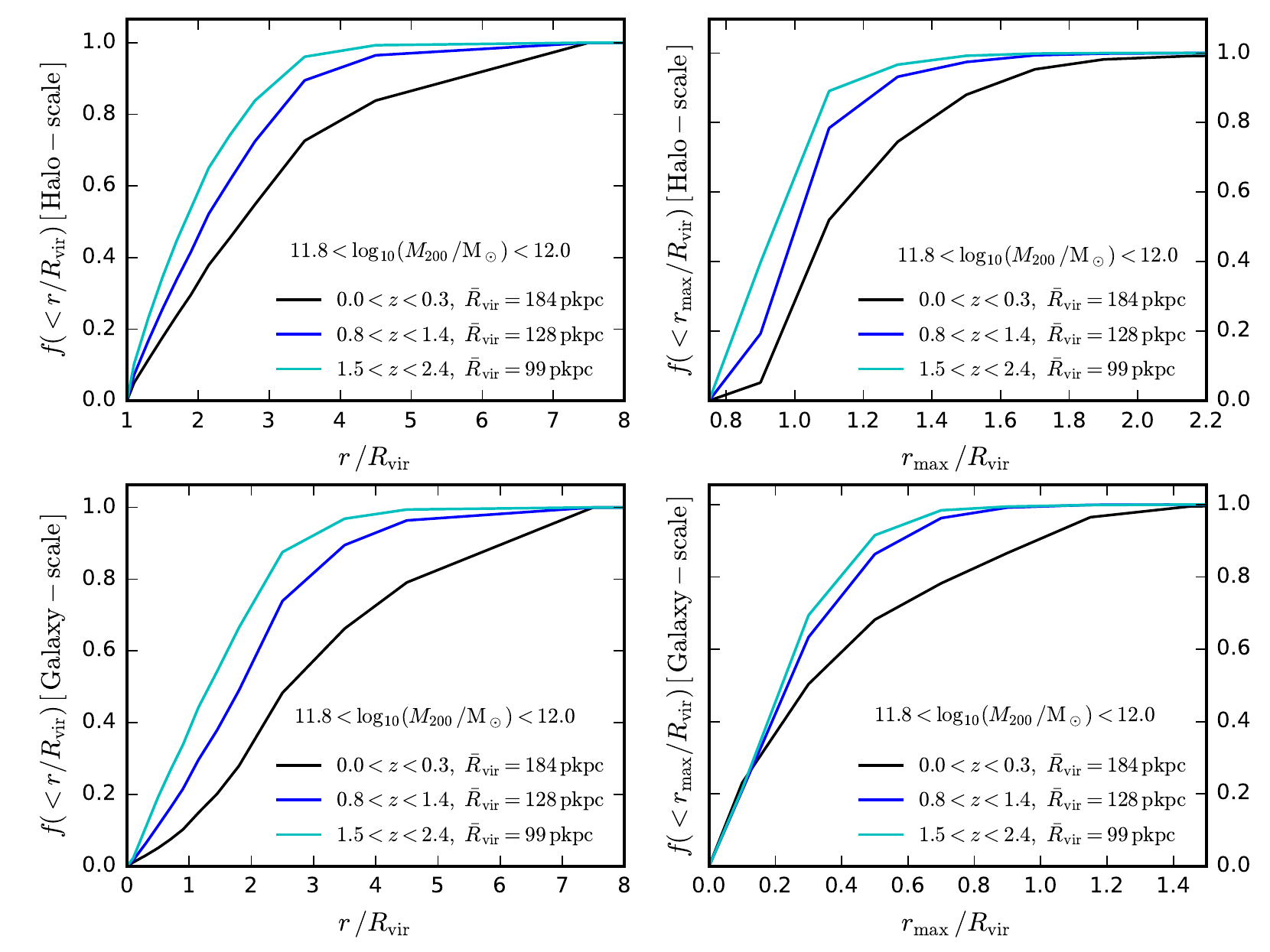}
\caption{The cumulative distributions of radius for gas ejected from haloes (top panels) and galaxies (bottom panels)
  for haloes with masses in the range $11.8 < \log_{10}(M_{200} \, / \mathrm{M_\odot}) < 12$ at the redshifts indicated.
  Left panels show the positions of ejected particles that have not returned to the halo/galaxy at the indicated redshift.
  Right panels show the maximum past radius of returning particles.
  In both cases radii are normalised by the median value of the halo virial radius 
  ($\bar{R}_{\mathrm{vir}}$) at each redshift interval plotted.
  Most of the gas ejected from galaxies resides beyond the virial radius (bottom-left panel),
  but most of the gas recycled onto galaxies was not ejected outside of the halo (bottom-right panel).
  Gas ejected from haloes (top-left panel) sits at roughly the same median position 
  (median value $\approx 2 R_{\mathrm{vir}}$ at $z \approx 0$) as gas ejected from galaxies (bottom-left panel).
}
\label{rmax_fig}
\end{center}
\end{figure*}

Finally, Fig.~\ref{rmax_fig} presents the cumulative distributions of distances reached
by ejected gas that is either currently (i.e. at the redshift indicated) returning to the halo (top-right) or
galaxy (bottom-right), or is currently residing outside the halo (top-left) or
galaxy (bottom-left). Distributions are plotted plotted for three
redshift bins with a fixed halo mass of $10^{11.8} < M_{200} \, /\mathrm{M_\odot} < 10^{12}$.
For returning gas (right panels), the distance plotted
is the maximum radius achieved, normalised to the current value of the halo
virial radius. For resident ejected gas (left panels), the distance plotted
is the current radius, normalised again to the current value of the halo
virial radius (for the halo from which the gas was ejected).

Most of the gas ejected from the ISM of galaxies resides beyond the virial
radius (bottom-left panel). For the plotted halo mass range, the median
radius of resident ejected gas is $2.6 \, R_{\mathrm{vir}}$ for $0<z<0.3$,
decreasing to $1.8 \, R_{\mathrm{vir}}$ for $0.8 < z < 1.4$, and to
$1.3 \, R_{\mathrm{vir}}$ for $1.5 < z < 2$.
Only $12 \, \%$ of the resident ejected gas is inside the virial radius
for $0<z<0.3$, though this fraction increases to $25 \, \%$ by
$0.8 < z < 1.4$, and to $37 \, \%$ by $1.5 < z < 2$.
The median distances of gas that has been ejected from haloes (irrespective
of having been in the ISM, top-left panel) are similar.
The maximum distance ever recorded for ejected gas (either from galaxies
or from haloes) is $\approx 1.3 \, \mathrm{pMpc}$ for the plotted
halo mass range.

For gas that is being recycled onto haloes (top-right) or galaxies (bottom-right),
the maximum distances achieved are much smaller. At the halo scale,
the median distance achieved is only $1.1 \, R_{\mathrm{vir}}$ for
$0<z<1.3$, and essentially all of the returning gas has
$r_{\mathrm{max}} < 2 \, R_{\mathrm{vir}}$.
Note that it is possible for returning gas to have $r_{\mathrm{max}} < R_{\mathrm{vir}}$;
this reflects the growth of the halo virial radius with time (the virial radius quoted
is the value just after gas has been recycled). Note also
that this gas must spend a significant amount of time
outside the halo, due to our adopted time-integrated velocity
cuts, see Section~\ref{tracking_sec}.
At the galaxy scale (bottom-right), returning gas has generally
never left the halo. The median maximum distance achieved
is $0.3 \, R_{\mathrm{vir}}$ at $0<z<0.3$, corresponding
to a fountain flow on scales of several tens of $\mathrm{pkpc}$.
Note again that there is expected to be another gaseous component
that participates in a smaller-scale galactic fountain 
(over shorter timescales), which is excluded from
our measurements (see Section~\ref{tracking_sec}).

\section{Literature Comparison}
\label{literature_section}

\subsection{First-time, recycled, and transferred inflow fractions}
\label{acc_mode_comp_sec}

\begin{figure}
\includegraphics[width=20pc]{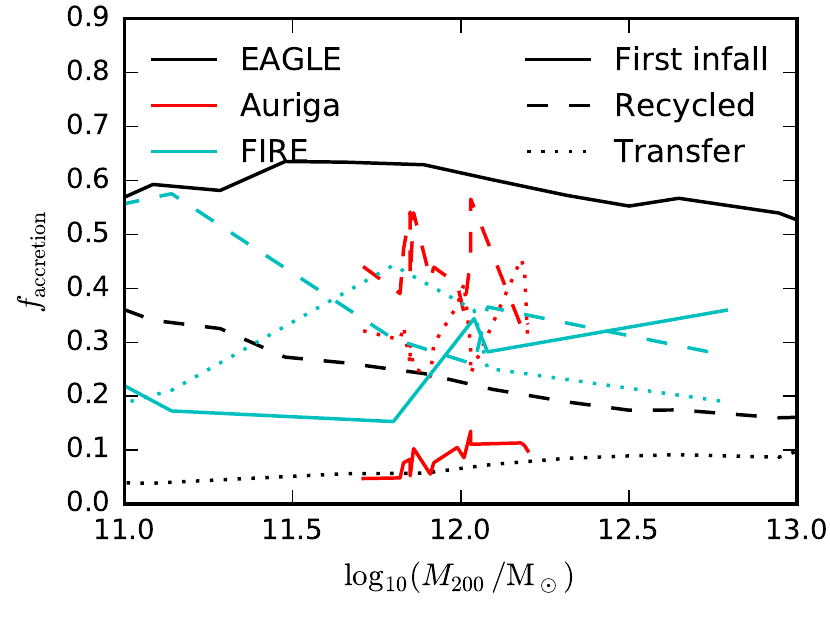}
\caption{The fraction of the final stellar mass contributed by different
  accretion channels at $z=0$, as a function of halo mass.
  Contributions are plotted for the ``First infall'' (solid), ''Recycled'' (dashed), and ``Transfer'' (dotted) channels ,
  as defined in the main text.
  We compare results from the reference \eagle simulation (black) with results from cosmological
  zoom-in simulation projects, including Auriga \protect \cite[red,][]{Grand19} and 
  FIRE \protect \cite[cyan,][]{AnglesAlcazar17}.
  Generally speaking, galaxies in \eagle are assembled mostly from first-infalling gas, whereas
  in Auriga galaxies are assembled mostly by gas that has been ejected and then recycled at least once.
  Over the same mass range as spanned by Auriga, FIRE galaxies present an intermediate scenario, 
  but with a stronger emphasis on transferred accretion.
}
\label{acc_mode_comp}
\end{figure}

We now consider how our results from \eagle compare to measurements presented in the literature
using other cosmological simulations, each of which uses a different implementation of
subgrid physics for star formation and feedback. We focus primarily here on the importance
and efficiency of wind recycling. Fig.~\ref{acc_mode_comp} compares the fractional contributions of 
first-time, recycled, and transferred gaseous inflow to the stellar mass of galaxies between
\eagle, and zoom-in simulations from the Auriga \cite[][]{Grand17} and FIRE simulation 
projects \cite[][]{Hopkins14}. For the cases of Auriga and FIRE, this fraction corresponds to
the fraction of the final stellar mass of $z=0$ galaxies. We use almost exactly the same
definition for \eagle, but in our case the fraction is
computed by integrating over the star formation history of all progenitor galaxies, and
is therefore weighted instead by the initial stellar mass of particles before stellar
mass loss (we do not expect this to bias our results significantly for galaxies at $z=0$). 
To be consistent with the other
studies, we use the main progenitor to define the dichotomy between  transferred and 
recycled gas for the comparison (see Section~\ref{tracking_sec}).

While we take these steps in order to make our measurements as comparable to the others
as is reasonably possible, it should be noted that there are still differences in
the methodologies used. Relative to our velocity cuts, \cite{AnglesAlcazar17} use
different velocity cuts to define ejected gas in FIRE (relying on the instanteous rather
than time-integrated velocity). We show in Appendix~\ref{ap_vcut} that the recycling is
fairly insensitive to the velocity cut used, and \cite{AnglesAlcazar17} arrive at the same conclusion after varying
their cuts, and so we do not expect this to affect the qualitative conclusions drawn
from the comparison.
The stellar feedback scheme of Auriga utilises
hydrodynamically decoupled wind particles, which enables the explicit identification of which
gas tracers in their simulations have been ejected from the ISM by stellar feedback,
removing the need for any velocity cuts. The characteristic maximum distance achieved
by recycled wind particles in Auriga is $\approx 20 \, \mathrm{kpc}$ (independent
of redshift), which is
uncomfortably close to the minimum allowed distance $\approx 15 \, \mathrm{kpc}$
implied by our velocity cuts for a Milky Way-mass halo at low redshift (though
our velocity cuts would correspond to smaller maximal distances for Milky Way
progenitor haloes at higher redshifts).
As mentioned previously, decreasing the velocity cut by a factor two (and
so halving the minimum allowed distance to be considered recycling) 
has little impact on our results (Appendix~\ref{ap_vcut})
\footnote{If we instead remove all velocity cuts, the contribution of recycling
would increase significantly (up to $50-60 \%$), but still not to the extent that Auriga and \eagle
would come into agreement for the contribution of first-infalling gas. 
From visual inspection of particle trajectories, we believe that the increase in
\eagle recycling rates for this case is more related
to particles fluctuating across the phase boundary defining the ISM, rather than 
reflecting genuine feedback-driven outflows however.}.
It is still important
to acknowledge that smaller-scale galactic fountain processes may occur
in reality however.

The explicit wind particle scheme of Auriga also allows \cite{Grand19} to
distinguish between gas that is removed from the ISM by feedback as opposed
to stripping processes (gravitational tides or ram pressure). Since neither
we nor \cite{AnglesAlcazar17} attempt to make this distinction, we group
the stripped and feedback-transferred components of \cite{Grand19} into a single
``transfer'' component for comparative purposes.

A final issue that could affect comparing \eagle with zoom-in simulations is
that \eagle utilises lower resolution. This is potentially important for the 
fraction of recycled and transferred accretion,
since \eagle may not resolve the ISM of very low-mass progenitor galaxies.
While very low-mass progenitor/satellite galaxies are negligible in terms of their 
stellar mass, and are naturally truncated by the preventative feedback associated with
photo-heating from the UVB, they are not necessarily negligible in terms of the
combined cumulative mass in outflows, since outflow rates per unit star formation 
increase strongly with decreasing stellar mass \cite[e.g.][]{Muratov15,Mitchell20}.
Using a higher-resolution simulation, we explore the effect of varying our
fiducial halo mass cut (above which we track ejected gas, and so count later recycling
and transfer) in Appendix~\ref{ap_smthr}, and find that the
contribution of recycled and transferred gas is reasonably well converged for
our fiducial mass cut.

Bearing these caveats in mind, Fig.~\ref{acc_mode_comp} shows
that \eagle predicts a significantly higher contribution from first-infalling
gas accretion, roughly $60 \%$, relative to recycling when compared to the Auriga simulations,
roughly $10 \%$.
Based on the various checks we have performed, we expect this conclusion to be 
qualitatively robust, and we interpret the difference as being caused by
the different implementations of stellar feedback between the simulations.
In Auriga, gas is ejected from galaxies by stellar feedback by explicit wind
particles that are temporarily decoupled from the hydrodynamical scheme,
and which typically recouple before reaching $10 \, \mathrm{kpc}$ in distance,
which is close to the average distance of $20 \, \mathrm{kpc}$ achieved by recycled particles \cite[][]{Grand19}.
In \cite{Mitchell20}, we compare outflow rates at different spatial scales
with the Illustris-TNG simulation \cite[][]{Nelson19}, which uses a very
similar implementation of stellar feedback as Auriga. There, we show that \eagle
drives outflows over much larger spatial scales at fixed mass (in the
mass range where stellar feedback dominates) than in Illustris-TNG.
This is consistent with the qualitative differences in the importance of
first-time versus recycled gaseous accretion between \eagle and Auriga, 
in the sense that recycling is less efficient in \eagle.

Roughly speaking, the FIRE simulations present a scenario that is intermediate
between the \eagle and Auriga cases (about $30 \%$ from first-time accretion), 
albeit with significant object-to-object scatter, and with a larger
contribution from the transferred component. Other studies find similarly
intermediate results between \eagle and Auriga: \cite{Christensen16} report 
that between $20 \%$ and $40 \%$ of gas accretion onto galaxies is recycled 
in their zoom-in simulations, and \cite{Ubler14} find $\approx 45 \%$ of the
gas accretion is first-time accretion. Note that these latter studies do not 
separate transferred accretion (and so we do not show them directly in 
Fig.~\ref{acc_mode_comp}).

\subsection{Recycling residency times}

\begin{figure}
\includegraphics[width=20pc]{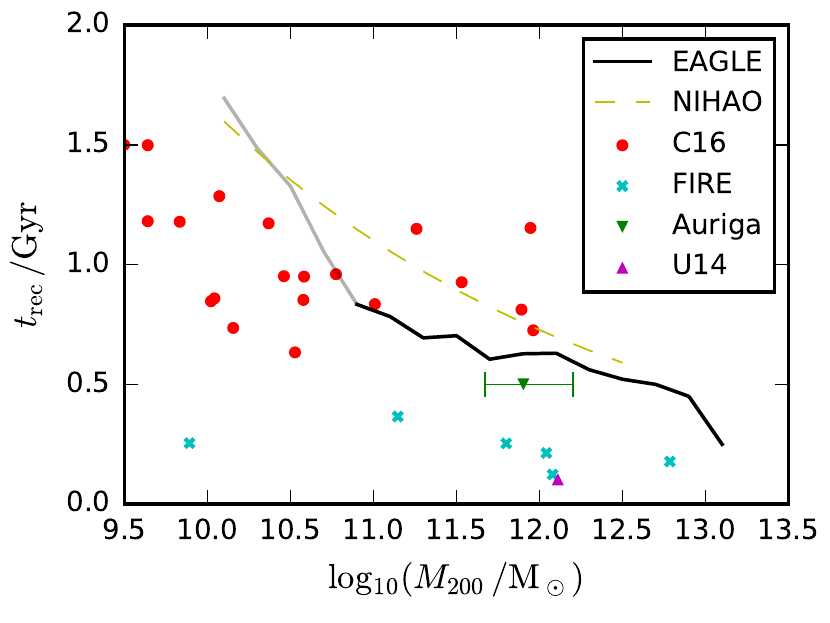}
\caption{The median residency time of returning gas after being ejected from the ISM of central galaxies 
  over the redshift range $0<z<3$.
  Note that this measurement excludes gas that does not return by $z=0$.
  We compare results from the reference \eagle simulation (black) with results from cosmological zoom-in simulation projects,
  including Auriga \protect \cite[green,][]{Grand19}, 
  FIRE \protect \cite[cyan,][]{AnglesAlcazar17},
  NIHAO \protect \cite[yellow,][]{Tollet19}, 
  and the zoom-in simulations of \protect \citet[][, C16, red]{Christensen16}
  and \protect \citet[][, U14, magenta]{Ubler14}.
  For \eagle, we indicate in grey the halo mass range for which galaxies contain fewer than $100$ stellar particles.
  With the exception of the measurements from FIRE and U14, 
  both of whom find short recycling timescales,
  there is good consensus among simulations that gas returns to galaxies over a timescale 
  between $0.5$ to $1 \, \mathrm{Gyr}$ for $M_{200} > 10^{11} \, \mathrm{M_\odot}$,
  and takes longer than $1 \, \mathrm{Gyr}$ to return at lower masses.
  We emphasise, however, that because this timescale excludes the contribution of non-returning particles,
  it is not a particularly useful way to characterise the efficiency of wind recycling.
}
\label{tret_comp}
\end{figure}

Most studies of wind recycling with cosmological simulations in the literature have characterised
recycling with the residency time for returning gas, i.e. the time between ejection and return.
 In some cases the median residency time is 
further taken as a measure of the efficiency of wind recycling, and is then compared to the 
parameterisations used in analytic and semi-analytic galaxy formation models. The flaw with this 
approach is that it neglects the contribution of gas that has not returned by $z=0$, and
so does not correctly characterise the recycling efficiency if the fraction of non-returning
gas is significant.

We compare the median residency time for gas that has been ejected from (and then returned to)
the ISM of galaxies between different cosmological simulations in Fig.~\ref{tret_comp}.
As before, the exact definitions of recycling vary from study to study, so quantitative
differences should not be over-interpreted.
For \eagle, FIRE, and the simulations of \cite{Christensen16}, the median residency time
is computed by averaging over all recycling events that occur over the range $0<z<3$.
For Auriga, \cite{Grand19} defines the residency time slightly differently as the time
between launching a wind particle and the time that the particle is either converted
into a star, or is re-launched as a wind particle. They compute this residency time
as a function of redshift, but since they find it is approximately $500 \, \mathrm{Myr}$
independent of redshift we simply overplot this value along with the other results. 
For the NIHAO zoom-in simulations, we take the best-fit relation quoted by \citet[][, their equation 11]{Tollet19}
for the mean residency time of gas that cycles from the ISM to a cool 
phase in the CGM (and back again), since this is the timescale that they use to compare with 
the efficiency of recycling in other models and simulations. 

Given the variety of definitions employed, Fig.~\ref{tret_comp} actually shows a
surprisingly good consistency between many of the different simulations.
Studies that sample a wider dynamic range in halo mass generally find that
the median residency time scales negatively with halo mass, ranging from
$\approx 500$ to $750 \, \mathrm{Myr}$ at $M_{200} \approx 10^{12} \, \mathrm{M_\odot}$.
The two apparent outliers are the results from FIRE and the single simulated halo
for which recycling times are presented in \cite{Ubler14}. These
two studies find significantly shorter recycling times of 
$\approx 200 \, \mathrm{Myr}$, and seemingly independent of halo mass in 
FIRE. We again caution that this could very well reflect differences
in the methodologies employed however.

\subsection{Recycling efficiencies}
\subsubsection{Halo scale}
\label{sam_comp_sec}

\begin{figure}
\includegraphics[width=20pc]{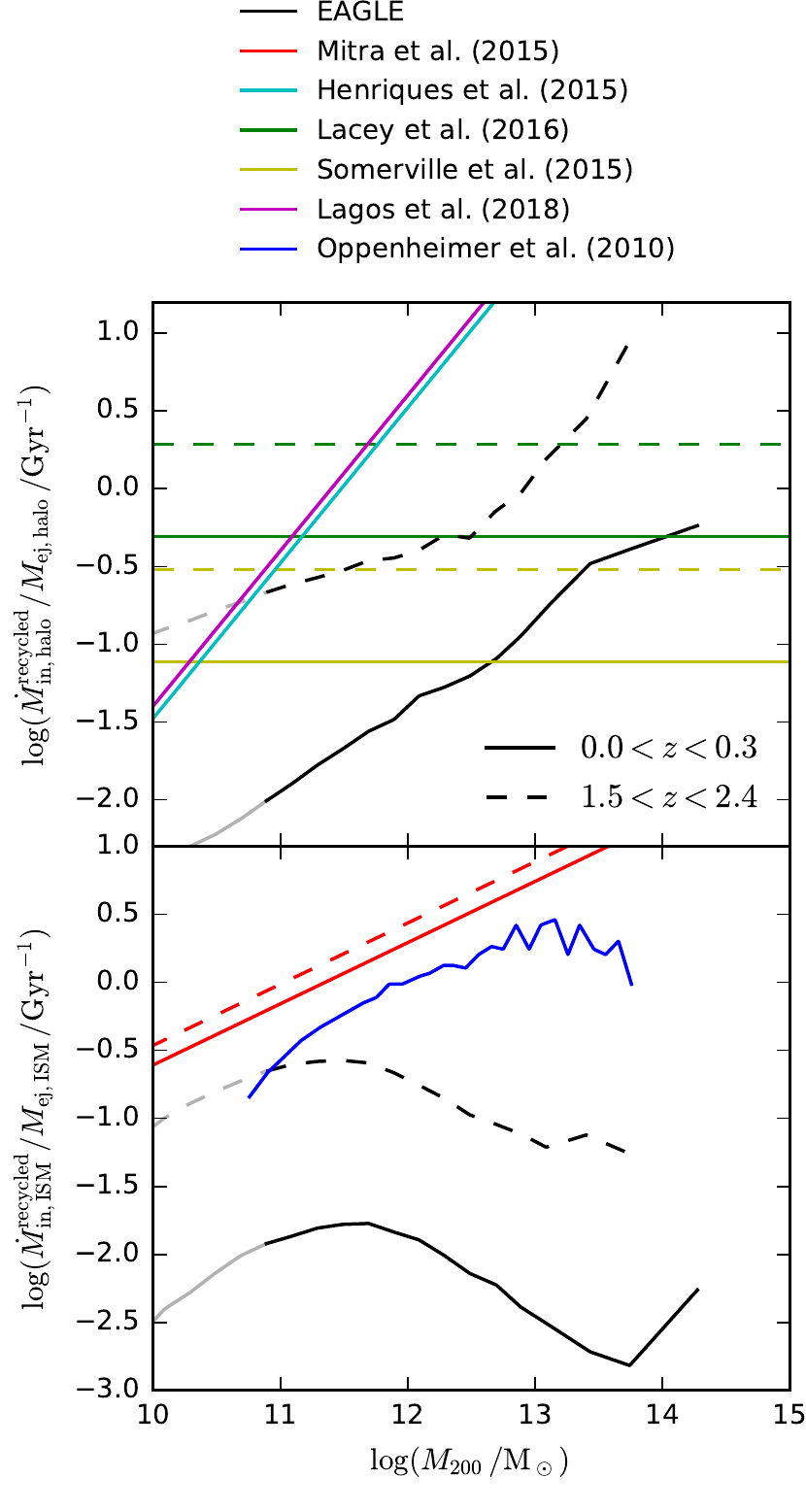}
\caption{The characteristic return efficiency for gas that is ejected from haloes (top panel),
  and from the ISM of galaxies (bottom panel), as a function of halo mass.
  We define the efficiency (per $\mathrm{Gyr}$ in this case) as 
  $\dot{M}_{\mathrm{in}}^{\mathrm{recycled}} \, / M_{\mathrm{ej}}$,
  where $\dot{M}_{\mathrm{in}}^{\mathrm{recycled}}$ is the inflow rate of
  recycled gas, and $M_{\mathrm{ej}}$ is the mass of the reservoir of ejected gas.
  At the halo scale, we compare results from the reference \eagle simulation (black) with 
  semi-analytic galaxy formation models, including 
  GALFORM \protect \cite[][]{Lacey16}, 
  L-GALAXIES \protect \cite[][]{Henriques15},
  SHARK \protect \cite[][]{Lagos18}, 
  and the Santa Cruz model \protect \cite[][]{Somerville15}.
  At the galaxy scale, we compare results from \eagle with the cosmological simulations
  of \protect \cite{Oppenheimer10}, and the idealised gas regulator model of \protect \cite{Mitra15}.
  Solid (dashed) lines show results at $z=0$ ($z=2$), with the exception of \protect \cite{Oppenheimer10},
  who measure the median time for gas ejected at $z \approx 1$ to return to the ISM (accounting for non-returning gas),
  as a function of halo mass at the time of ejection.
  At the halo scale, the efficiency of gas recycling in \eagle varies strongly with redshift
  (fairly consistent with models that assume the efficiency scales with the halo dynamical time),
  and with a halo mass dependence  that is intermediate between the two often considered cases of
  no dependence, and linear scaling with halo mass.
  At the galaxy scale, \eagle finds longer timescales than \protect \cite{Oppenheimer10}
  and \protect \cite{Mitra15}, particular at high halo masses where the efficiency drops in \eagle
  due (at least in part) to the presence of AGN feedback, which was not included in the simulations
  of \protect \cite{Oppenheimer10}.
  Overall, the differences between the different models are very large.
}
\label{rec_eff_comp}
\end{figure}

The average residency time (the time between ejection and return) is a poor measure of the efficiency
of wind recycling. This is underlined strongly for the specific case of recycling in 
the \eagle simulations, since the median residency time for returning gas 
is $\approx 0.5$ to $1 \, \mathrm{Gyr}$ (Fig.~\ref{tret_comp}), but the characteristic return time
for for all ejected gas to return ranges from $3$ to hundreds of $\mathrm{Gyr}$,
depending on the halo mass and redshift (Figs.~\ref{rec_eff_fig} and \ref{rec_eff_comp}). 

As discussed in Section~\ref{rec_sec}, we instead measure the efficiency of wind recycling
in \eagle by taking the ratio of the inflow rate of returning gas to the mass of resident
ejected gas\footnote{Meaning ejected gas that remains outside the ISM for galaxy-scale
recycling, or remains outside the virial radius for halo-scale recycling}. This is also 
the definition of recycling efficiency that is used in most
semi-analytic galaxy formation models, and we now consider a comparison between the two.
Semi-analytic models generally model recycling explicitly at the halo
scale only, with ``returning'' gas in this case referring to the rate of return of
gas back onto the normal ``halo-gas'' reservoir of circum-galactic gas that resides outside
the ISM but still within the halo virial radius.
It is usually assumed that the gas ejected beyond the virial radius is 
powered by stellar feedback, with AGN feedback instead utilised as a mechanism
to prevent baryons within the virial radius from reaching the ISM 
\cite[though see, for example,][for alternative schemes]{Bower08}.

The details then vary slightly from model to model. The \galform model 
\cite[as of][]{Lacey16} makes the assumption that the gas ejected from the ISM of galaxies 
is also ejected beyond the halo virial radius. 
The Santa Cruz model \cite[as of][]{Somerville08,Somerville15} 
assume that a fraction of the gas ejected
from the ISM is immediately reincorporated into the standard halo-gas reservoir,
and that a fraction is ejected beyond the halo virial radius (their recycling
efficiency then refers to the latter component). The L-Galaxies
model of \cite{Henriques15} and SHARK model of \cite{Lagos18} make a similar assumption, but also allow for
the ejection of halo gas that was not previously part of the ISM. 
In the L-Galaxies model, the ``ejected''
gas is conceptually considered to be a combination of gas that
is genuinely ejected beyond the virial radius, and of gas that is spatially
located within the halo but which is too hot to undergo standard infall
from the CGM (this has the disadvantage of leaving the total halo baryon content 
as a not clearly defined quantity).

Broadly speaking, the definitions of recycling efficiency used in these models are all
at least roughly equivalent to our measurements of halo-scale
recycling in \eagle, and we show the comparison between the two in the top panel
of Fig.~\ref{rec_eff_comp}. Tellingly, the efficiency of halo-scale
recycling can vary by over two orders of magnitude between different
models and \eagle at a given halo mass and redshift, despite the fact 
that each was calibrated to reproduce the same basic galaxy population 
diagnostics (such as the galaxy stellar mass function or luminosity function). 
In \cite{Mitchell20}, a comparison between \eagle
and the same models for galaxy and halo-scale outflows shows differences
of comparable order. As discussed in that work, this largely reflects
the underlying degeneracy in galaxy formation that the integrated 
stellar properties of galaxies (that are usually used as constraints for
theoretical models and simulations) are only
sensitive to the \emph{net} inflow rates of gas onto galaxies, and
not to the relative breakdown between first-time inflows, outflows,
and wind recycling.

The top panel of Fig.~\ref{rec_eff_comp} does show some commonality however.
The \galform and Santa Cruz models assume that the efficiency of
halo-scale wind recycling scales inversely with the age of the Universe,
which is similar to the evolution in \eagle at fixed halo mass
(as demonstrated explicitly in Fig.~\ref{rec_eff_fig}), albeit
the redshift evolution is stronger in \eagle. 
Conversely, both the \cite{Henriques15} L-galaxies model and 
the (fiducial) SHARK model of \cite{Lagos18} assume that the 
recycling efficiency depends linearly and positively with halo mass,
which is again similar to \eagle at fixed redshift, though
in this case the halo mass dependence is weaker in \eagle.

\subsubsection{Galaxy scale}

The bottom panel of Fig.~\ref{rec_eff_comp} compares
the efficiency of galaxy-scale wind recycling between \eagle,
the cosmological simulation of \citet[][, we compare
to their preferred ``VZW'' momentum scaling wind model]{Oppenheimer10}, and
the best-fit ``equilibrium'' model presented in \cite{Mitra15}.
Both \cite{Oppenheimer10} and \cite{Mitra15} measure wind
recycling at the galaxy scale (i.e. gas that is ejected from
and returns to the ISM of galaxies). As with \cite{Grand19},
\cite{Oppenheimer10} characterise wind recycling by measuring
the difference between the time wind particles are launched, and the time wind
particles either form a star or are re-launched.
Unlike other the analyses presented for other cosmological
simulations however, \cite{Oppenheimer10} include the contribution
of non-returning gas particles by setting their residency
time to the age of the Universe, and only quote a median
if it is less than this value. While not precisely the same
as the definition used for \eagle \cite[see Section~\ref{rec_sec},
which is formally equivalent
to the definition used in][]{Mitra15}, the two quantities
are close enough that we believe that the comparison is
meaningful. A final caveat however is that
\cite{Oppenheimer10} measure the efficiency of recycling
for gas that is ejected at $z=1$, and that returns
(or not) over the range $0<z<1$, as a function of
the halo mass at the time of ejection. This does not
map straightforwardly onto our measurements in \eagle
\cite[or to][]{Mitra15},
but we mitigate this by showing our recycling efficiencies
at $z=2$ and at $z=0$, which should bracket the
range used in \cite{Oppenheimer10}.

The bottom panel of Fig.~\ref{rec_eff_comp} shows
that galaxy-scale recycling is apparently more efficient in the
simulation of \cite{Oppenheimer10} than in \eagle, although
our $z=2$ results overlap at $M_{200} \approx 10^{11} \, \mathrm{M_\odot}$.
Unlike in \eagle, where the efficiency of galaxy-scale
wind recycling peaks at $M_{200} \approx 10^{12} \, \mathrm{M_\odot}$ (and
declines strongly at higher masses), the
efficiency of wind recycling continues to rise monotonically
with halo mass in the simulation of \cite{Oppenheimer10}. This
has the consequence that almost all of the gas accretion 
onto massive galaxies is recycled in this simulation.
Their simulation also significantly over-estimates the
abundance of massive galaxies, and both of these properties
presumably reflect the lack of any form of AGN feedback
in their simulation
\cite[although see][for discussion of additional caveats to their
stellar feedback model]{Huang20}.

The comparison with the best-fit equilibrium model of \cite{Mitra15}
is also worthy of discussion. \cite{Mitra15} constrain their
model against empirically-derived constraints on the relationship
between galaxy stellar mass and halo mass, the relationship between
galaxy star formation rate and stellar mass, and the relationship
between ISM metallicity and stellar mass, all as a function
of redshift. Given their adopted parameterisations
for preventative feedback, outflows, and recycling, they find
a unique solution that fits the data within their parameter
space. Furthermore, they report that using only two of the
three observational constraints is sufficient to constrain the model.
It is therefore interesting to note the greater than order
of magnitude difference in recycling efficiency between
their best-fit model and \eagle at $z=0$, especially given 
that \eagle was calibrated to produce very similar stellar masses. 
\eagle does systematically underpredict
galaxy star formation rates for actively star-forming
galaxies at a given stellar mass by about $0.3 \, \mathrm{dex}$,
but otherwise the simulation predicts star formation rates
that evolve in time in a way that is broadly consistent with
observations \cite[][]{Furlong15}. 
In addition, while the reference \eagle model (at fiducial resolution) does not
agree well with the observed galaxy mass-metallicity relationship
for $M_\star < 10^{10} \, \mathrm{M_\odot}$,
reasonably good agreement is seen
for the higher resolution Recal simulation down to 
$M_\star \sim 10^9 \, \mathrm{M_\odot}$ \cite[][]{Schaye15}.
The Recal simulation has qualitatively very similar recycling efficiency
to that seen here (see Appendix~\ref{ap_num_con}), at least relative to the discrepancy
between \eagle and the model of \cite{Mitra15}.

Putting this together, this reinforces the idea that the
stellar properties of galaxies (and possibly even the mass-metallicity
relationship) do not strongly constrain the overall network
of first-time inflows, outflows and recycling that regulate
galaxy evolution. To make progress with simplified analytic
and semi-analytic models, we suggest that appropriately
defined measurements need to be performed for different 
hydrodynamical simulations with different implementations
of feedback physics, and that the parameterisations adopted
in simple models should then be made flexible enough to
match the range in inflow/outflow/recycling efficiencies
measured from simulations. With this in place, we suggest
that a parameter space search using similar observational
constraints to those employed in \cite{Mitra15} may yield
more robust results, albeit with the (likely) conclusion
that conventional observational diagnostics of the
galaxy population are indeed not sufficiently constraining.
The observations that presumably would provide
more constraining power, primarily the distribution
of ejected metals as a function of distance from the
host galaxy, are less readily modelled in simplified
analytic or semi-analytic models, but again this could perhaps
be mitigated by taking results from different
hydrodynamical simulations in the literature to provide
a physical prior on which distributions would be reasonable,
and to enable a self-consistent link between the assumed
mass and energy fluxes of outflows, and the resulting
spatial distribution of extra-galactic metals.

\subsection{Inflows with super-Lagrangian refinement in the CGM}

\cite{Vandevoort19}, \cite{Peeples19}, \cite{Hummels19} 
and \cite{Suresh19} have recently explored the effects of using
non-Lagrangian criteria to increase the numerical resolution inside
the CGM, using cosmological zoom-in simulations. The relative
lack of numerical resolution inside the CGM of cosmological 
simulations has been a longstanding point of discussion, and
the afore-mentioned studies show that tracers of dense gas in
the CGM do change when resolution of low-density CGM phases
is increased.

The impact of this on inflowing gas fluxes onto galaxies is
currently unclear, although presumably fairly minor since the
properties of the host galaxy are not reported to change significantly
when the CGM resolution is changed \cite[][]{Peeples19}.
The detailed trajectories and relative importance of different
gas accretion channels could still be affected however.
If thermal instabilities do lead to significant
conversion of mass from a hot thermal wind into cold dense
clumps, this would presumably have a significant impact
on subsequent gas recycling. The relatively dense phases
of gas that are stripped from satellites could also be 
conceivably affected, perhaps altering the importance of
the transfer and merger components discussed in this work.
These questions look set to inspire continued work
using zoom-in simulations in the coming years.

\section{Summary}
\label{summary_section}

We have measured gaseous inflow rates onto galaxies
and their associated dark matter haloes in the \eagle simulations.
By tracking particles after they are ejected from galaxies and/or
haloes, we quantify the relative importance and efficiencies
of first time gaseous-infall, wind recycling, and the transfer
of ejected gas between independent galaxies and haloes.
For wind recycling, we select only gas that first outflows
with a time-integrated radial velocity greater than one
quarter of the halo maximum circular velocity, over at least
one quarter of a halo dynamical time, corresponding to a
radial displacement of $\approx 15 \mathrm{kpc}$ for a Milky Way-mass 
halo at low redshift. Our measurements focus
therefore on recycling of gas that has moved outwards over spatial
and temporal scales that are significant relative to the associated 
scales of the dark matter halo, and do not account for any
galactic fountains that occur on smaller scales at 
the disk-halo interface.

Qualitatively consistent with earlier results from the
OverWhelmingly Large Simulations project \cite[OWLS,][]{VanDeVoort11},
we find that for gas accretion onto the ISM of galaxies, 
inflow rates per unit halo mass clearly peak for a halo mass 
of $\sim 10^{12} \, \mathrm{M_\odot}$
(Fig.~\ref{gas_accretion_figure}), 
which helps to explain the empirically inferred relationship
between galaxy stellar mass and halo mass 
\cite[e.g.][]{Behroozi10,Moster10}, and therefore the shape of the observed 
galaxy stellar mass function.

Gaseous accretion rates onto haloes are reduced
relative to (scaled) dark matter accretion rates for haloes
with $M_{200} < 10^{13} \, \mathrm{M_\odot}$.
As demonstrated in \cite{Wright20}, the
reduction of gas accretion (relative to dark matter accretion) 
at the virial radius is connected
primarily to the implementation of feedback processes
in \eagle, which is demonstrated by comparing halo-scale accretion
rates of first-time and recycled infall between the fiducial
\eagle simulations and simulation variations with
feedback and/or radiative cooling processes removed. Here, we show that instantaneous 
first-time gaseous infall is slightly suppressed relative to 
first-time infalling dark matter for 
$M_{200} < 10^{11} \, \mathrm{M_\odot}$, but actually exceeds 
dark matter accretion at higher halo masses (Fig.~\ref{prev}). 
Considering instead the time-integrated mass of first-time
infall at the viral radius (Fig.~\ref{prev_z0}), we show that the
integrated gas accretion onto haloes is slightly suppressed for
$M_{200} < 10^{12} \, \mathrm{M_\odot}$ (by up to $0.4 \, \mathrm{dex}$), 
and is slightly enhanced at higher halo masses (by up to $0.2 \, \mathrm{dex}$).
At the transition mass of $M_{200} = 10^{12} \, \mathrm{M_\odot}$,
we show in Fig.~\ref{prev_evo} that first-time gas accretion at the virial
radius is not ``prevented'' at the virial radius by $z=0$, but is
rather delayed from the peak epoch of specific halo growth (at
$z \approx 4$), to slightly later times ($z \approx 3$). 
This effect may partly explain why
observed galaxy growth does not appear to closely trace the time evolution
of predicted halo mass growth \cite[][]{Daddi07,Dekel14,Mitchell14}.

Splitting gas accretion onto galaxies and haloes between 
first-time infall, recycled accretion, transfer between 
independent galaxies/haloes, and mergers, we find that first-time
infall usually provides the largest single contribution
(Fig.~\ref{return_contr}). This differs from some 
other simulations with different implementations of 
stellar feedback \cite[][]{AnglesAlcazar17,Grand19}, who
find significantly higher contributions from recycled
or transferred gas (Fig.~\ref{acc_mode_comp}).
The results from \eagle follow from our study of outflows,
presented in \cite{Mitchell20}, where we show that,
relative to other simulations, outflows in \eagle
are driven out over larger scales, and with mass fluxes
that increase with radius, such that more gas is being
ejected from haloes than from the ISM of galaxies. In this
picture, the natural expectation is that recycling
is less efficient as a mechanism to bring gas back
to galaxies. We note that at least for low-mass haloes,
outflows must reach far beyond the halo virial radius
to account for the high rate of incidence of weak metal
lines in blind quasar absorption surveys \cite[e.g.][]{Booth12}.

We measure the efficiency of first-time infall from the
CGM onto the ISM (Fig.~\ref{infall_figure}), defined
as the ratio between cosmic time and the characteristic
timescale for the CGM to accrete onto the galaxy.
This efficiency is insensitive to
halo mass for $M_{200} < 10^{12} \, \mathrm{M_\odot}$,
and drops sharply at higher halo masses. This
is qualitatively consistent with the traditional
picture for galaxy formation, in which gas infall
within low-mass haloes is limited by gravitational
timescales (which are scale free), and by radiative
cooling timescales (which are increased by AGN feedback)
in higher-mass haloes 
\cite[see also][]{Correa18b,Davies20}. At
fixed halo mass, infall efficiencies decrease 
with time, which for 
$M_{200} < 10^{12} \, \mathrm{M_\odot}$ implies
that there is more than just a basic dependence
on the radial gravitational freefall time.

We also measure the efficiency of wind recycling
at both galaxy (recycling of gas leaving/rejoining the ISM) and halo
(recycling of gas at the virial radius) scales (Fig.~\ref{rec_eff_fig}),
with the efficiency defined as the ratio between cosmic time and
the characteristic timescale for ejected gas
to return, accounting for gas that never returns.
The efficiency of halo-scale wind recycling
evolves at fixed mass, decreasing
at lower redshifts.
The efficiency of halo-scale wind
recycling increases with halo mass, but
with a power law exponent ($\approx 0.6$) that
is smaller than the value of unity
adopted by several recent semi-analytic
galaxy formation models \cite[][]{Henriques13,Hirschmann16,Lagos18}.

Wind recycling onto the ISM of galaxies is generally
less efficient than wind recycling onto haloes
(Fig.~\ref{rec_eff_fig}), and (when considering all of the gas
ejected from galaxies) clearly peaks at a characteristic
halo mass of $M_{200} \sim 10^{12} \, \mathrm{M_\odot}$,
reflecting the peak in total galaxy accretion rates
at this mass. Wind recycling therefore plays an
active role in shaping the characteristic Schecter-function-like
shape of the galaxy stellar mass function in \eagle.
If we consider the efficiency of wind recycling
for only the subset of gas that has not escaped the halo
(Fig.~\ref{rec_eff_fig_cgm}), we find that the efficiency of
CGM-scale wind recycling is actually higher
than the efficiency of first-time infall from
the CGM, and declines strongly with increasing halo mass.

At $z=0$, most of the gas ejected
from galaxies resides between $2$ and
$4$ times the halo virial radius (Fig.~\ref{rmax_fig}).
Gas that returns to galaxies generally
has not left the halo however, with half
of the returning gas at $z=0$ having 
reached only a quarter of the halo virial radius
before falling back.

By comparing the efficiency of wind recycling
between \eagle and semi-analytic models,
we show differences between models and
\eagle that span orders of magnitude at
fixed halo mass and redshift (Fig.~\ref{rec_eff_comp}),
highly reminiscent of a similar comparison of 
outflow rates presented in \cite{Mitchell20}. 
This strongly emphasises the point 
that the ``baryon cycle'' (meaning the overall
network of inflows, outflows and recycling)
is deeply degenerate, with conventional
extra-galactic observational constraints
only constraining the \emph{net} inflow of
gas onto galaxies.

With this degeneracy in mind, it would be
very timely to review if cosmological
simulations (each with different implementations
of uncertain star formation and feedback
processes) yield similarly discrepant recycling
efficiencies. Unfortunately, the majority of 
analyses of wind recycling
in state-of-the-art cosmological simulations
have not presented measurements of the efficiency
of wind recycling in a way that accounts for
non-returning gas (which is an essential component
in the \eagle simulations), and we
strongly encourage future analyses of other
simulations to present such measurements.

\section*{Acknowledgements}

We would like to thank Dr. John Helly for producing and sharing the halo
merger trees that form the backbone of our analysis. We thank
Claudia Lagos and Ruby Wright for useful discussions.

This work used the DiRAC@Durham facility managed by the Institute for
Computational Cosmology on behalf of the STFC DiRAC HPC Facility
(www.dirac.ac.uk). The equipment was funded by BEIS capital funding
via STFC capital grants ST/K00042X/1, ST/P002293/1, ST/R002371/1 and
ST/S002502/1, Durham University and STFC operations grant
ST/R000832/1. DiRAC is part of the National e-Infrastructure.

This work was supported by Vici grant 639.043.409 from
the Netherlands Organisation for Scientific Research (NWO).
RGB acknowledges support from the Durham consolidated grant:
ST/P000541/1.

\section*{Data availability}

The data underlying this article will be shared on reasonable request to the corresponding author.
Raw particle data and merger trees for the \eagle simulations have been publicly released \cite[][]{McAlpine16}.

\bibliographystyle{mn2e}
\bibliography{bibliography}

\appendix
\section{Convergence}
\label{ap_num_convergence}

\subsection{Resolution convergence}
\label{ap_num_con}

\begin{figure*}
\begin{center}
\includegraphics[width=40pc]{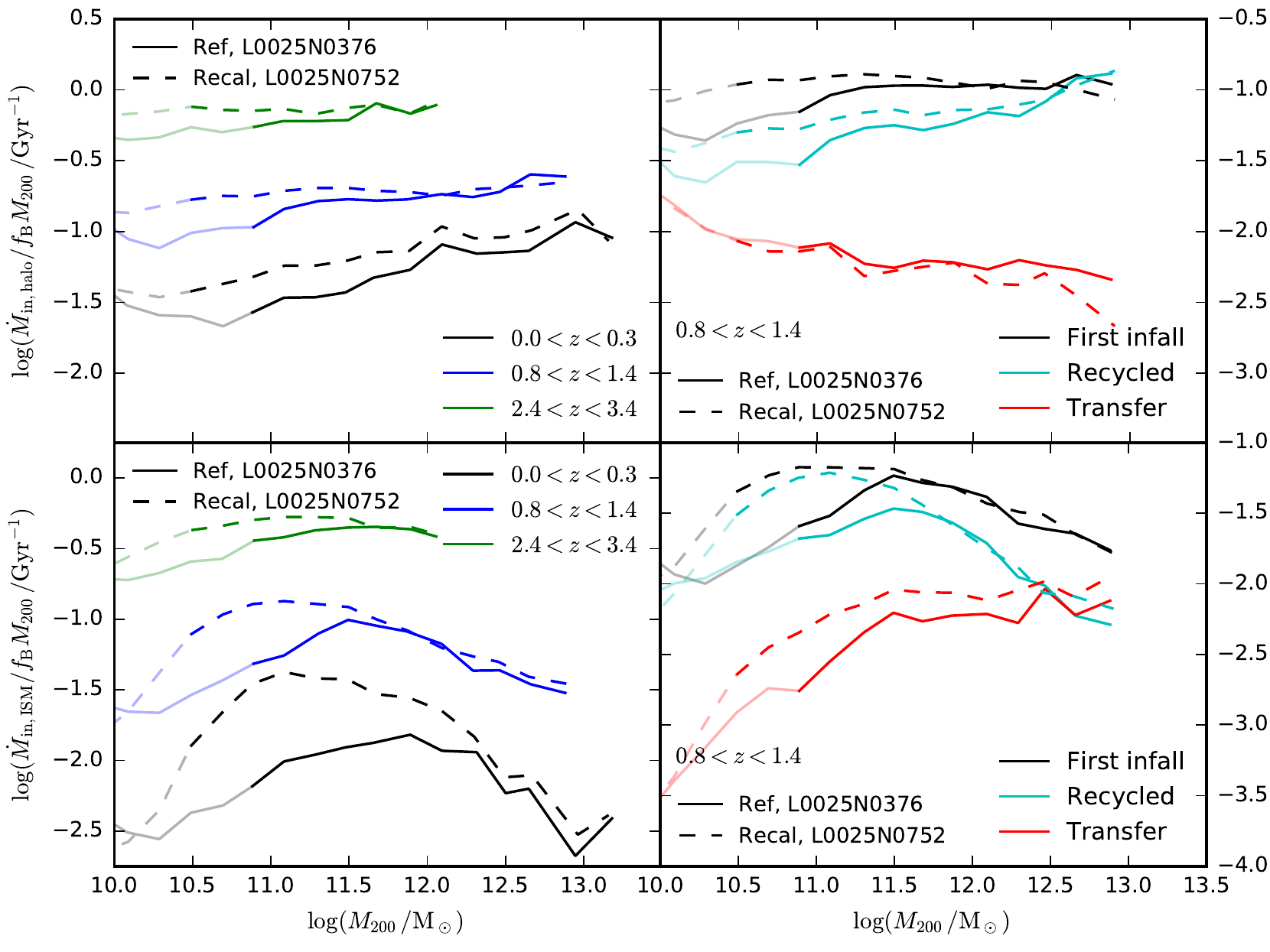}
\caption{A comparison of gaseous inflow rates between a
  standard-resolution \eagle simulation of a $(25 \, \mathrm{Mpc})^3$ volume with reference-model parameters 
  (Ref, L0025N0376, where the naming convention is L $+$ box length $+$ N $+$ cube root of the number of particles),
  and a recalibrated higher-resolution (eight times higher mass resolution) re-simulation of the same volume (Recal, L0025N0752).
  Top (bottom) panels show inflow rates onto haloes (galaxies).
  The left panels show the total smooth inflow rate for three redshift ranges, as labelled.
  The right panels show the separate contributions from first-infalling gas, recycled gas, and transferred gas, 
  over the interval $0.8 < z < 1.4$.
  Transparent lines indicate the range where there are fewer than $100$ stellar particles per galaxy.
  Inflow rates are reasonably well converged for $M_{200} > 10^{11.5} \, \mathrm{M_\odot}$,
  but are higher in the higher-resolution Recal simulation for 
  $M_{200} < 10^{11.5} \, \mathrm{M_\odot}$, 
  most notably on galaxy scales at $M_{200} \approx 10^{11} \, \mathrm{M_\odot}$ at $z \approx 0$.
  This also applies to each of the individual inflow channels (right panels), with the exception of halo gas transfer.
}
\label{recal_comp}
\end{center}
\end{figure*}

Fig.~\ref{recal_comp} compares gas accretion rates between the Reference \eagle
(at fiducial \eagle resolution) and Recal \eagle models (at eight times 
higher mass resolution). We use a common $(25 \, \mathrm{Mpc})^3$ volume for both models.
The left panels of Fig.~\ref{recal_comp} show that inflow rates are reasonably
converged between the two resolutions in higher-mass haloes, $M_{200} > 10^{12} \, \mathrm{M_\odot}$
at $z \approx 0$ and $M_{200} > 10^{11} \, \mathrm{M_\odot}$ at $z \approx 3$,
but are not well converged at lower halo masses, especially at the galaxy scale
(bottom-left panel) for which inflow rates can be up to $\approx 0.5 \, \mathrm{dex}$
higher in the Recal model (for $M_{200} = 10^{11} \, \mathrm{M_\odot}$ at $z=0$).
The higher inflow rates seen in the Recal model also apply to each of the
separated contributions from first-time, recycled, and transferred
accretion (right panels). The exception is transferred accretion
at the halo scale, which is well converged between the two 
simulations.

Qualitatively, our results remain similar between the Reference and
Recal models. The most clear and important quantitative difference is the characteristic
halo mass for which galaxy-scale inflow rates peak, which is lower (and the
peak broader) in the Recal model, at $M_{200} \approx 10^{11} \, \mathrm{M_\odot}$.
We intend to explore the implications of this difference for the relationship
between galaxy stellar mass and halo mass in future work
\cite[the stellar mass to halo mass ratio does still peak at 
$M_{200} \approx 10^{12} \, \mathrm{M_\odot}$ in the Recal model,][]{Schaye15}.

\subsection{Temporal convergence}

\begin{figure}
\includegraphics[width=20pc]{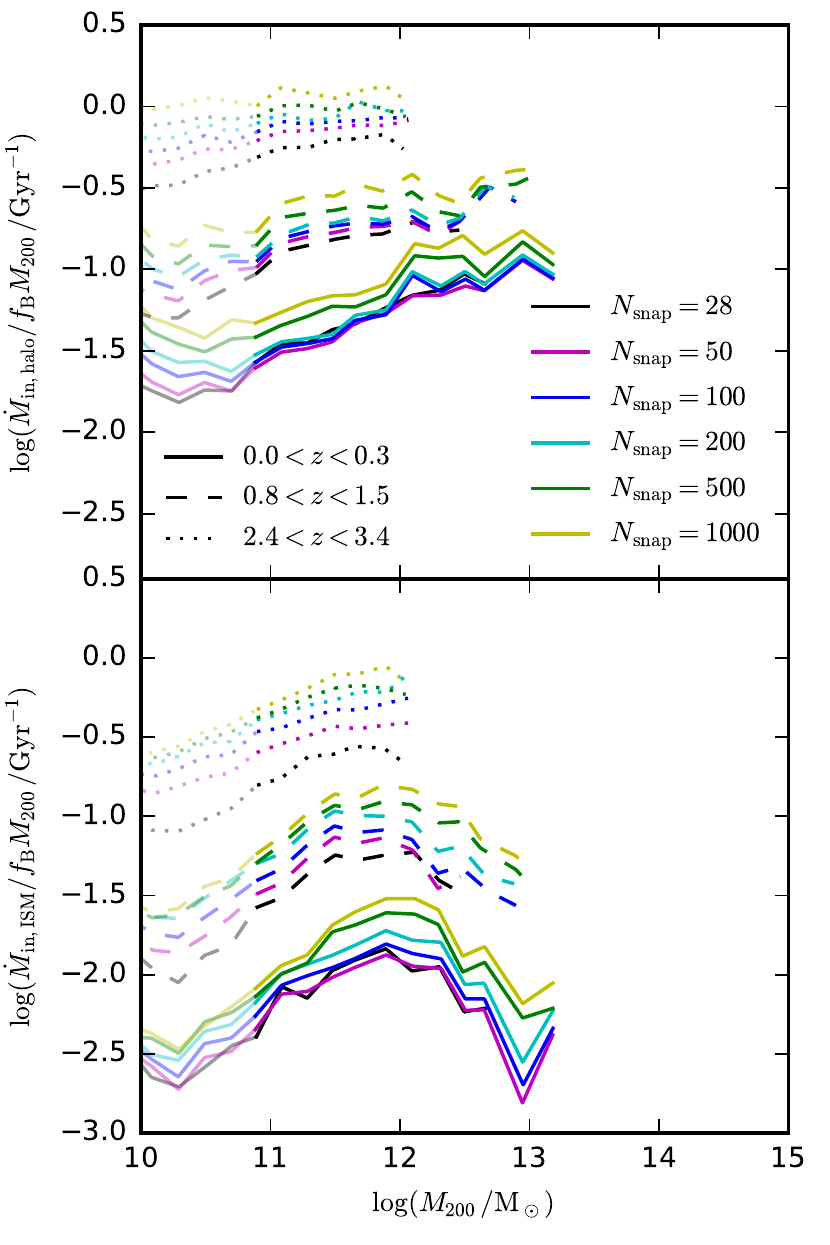}
\caption{The dependence of inflow rates on the snapshot cadence.
  Top (bottom) panels show total gaseous accretion rates onto haloes (onto galaxies), plotted as a function of halo mass.
  Different lines correspond to different numbers of simulation snapshots used to perform the analysis,
  with $200$ being the reference number used in this study
  \protect \cite[see appendix A1 in][for the precise time spacing]{Mitchell20}.
  Here, we use a smaller $(25 \, \mathrm{Mpc})^3$ volume simulation, for which a higher snapshot cadence is available.
  Transparent lines indicate the range where there are fewer than $100$ stellar particles per galaxy.
  Temporal convergence is reasonably good at the ISM scale (but not excellent at higher masses at lower redshifts).
  Convergence is less convincing at the halo scale, but only affects the normalisation of our results, at the level
  of tens of percent.
  }
\label{time_convergence_plot}
\end{figure}

We use $200$ discrete simulation outputs (referred to as snapshots) to track 
the movements of particles in post-processing from our simulations.
The time sampling of these snapshots is shown in appendix A1 of \cite{Mitchell20}.
This sampling affects our results in the sense that insufficient time
resolution will cause us to miss inflowing particles that are then
ejected within a timescale that is smaller than the separation between snapshots. 
In \cite{Mitchell20}, we showed that outflow rates are fairly well (but not fully) 
converged for our fiducial snapshot grid.

Fig.~\ref{time_convergence_plot} shows the corresponding picture for total
inflow rates onto haloes (top) and galaxies (bottom), using a $(25 \, \mathrm{Mpc})^3$ 
volume for which five times more snapshots are available. As with outflows,
we find that galaxy-scale inflow rates converge fairly well for
$M_{200} < 10^{12} \, \mathrm{M_\odot}$, but convergence is less
good at higher-halo masses, particularly at low redshift.
Inflow rates are less well converged at the halo scale
for low-mass galaxies, and increase by roughly $50 \%$ after
increasing the number of snapshots by a factor $5$ relative to our
fiducial spacing. If we instead consider the fractional contribution 
of different accretion channels to the total inflow rate (not shown),
the fractional contributions are well converged at our fiducial
snapshot cadence at the halo scale, and for $M_{200} \geq 10^{12} \, \mathrm{M_\odot}$
at the galaxy scale. At the galaxy scale, the fractional contribution of (for example) recycled
gas increases by
$\approx 30 \%$ at $z \sim 0$ if we increase the snapshot cadence
by a factor five for $M_{200} \sim 10^{11} \, \mathrm{M_\odot}$.

Quantitatively our results are therefore affected by the available
snapshot cadence, although qualitatively our results are unaffected
\cite[consistent with conclusions of a similar check by][]{Vandevoort17}.
This guides our choice to use as many simulation snapshots as possible,
which is $200$ for the flagship simulation.

\section{Methodology details}
\label{ap_methods}

\subsection{Varying the definition of recycling versus transferred accretion}
\label{ap_mp}

\begin{figure} 
\includegraphics[width=20pc]{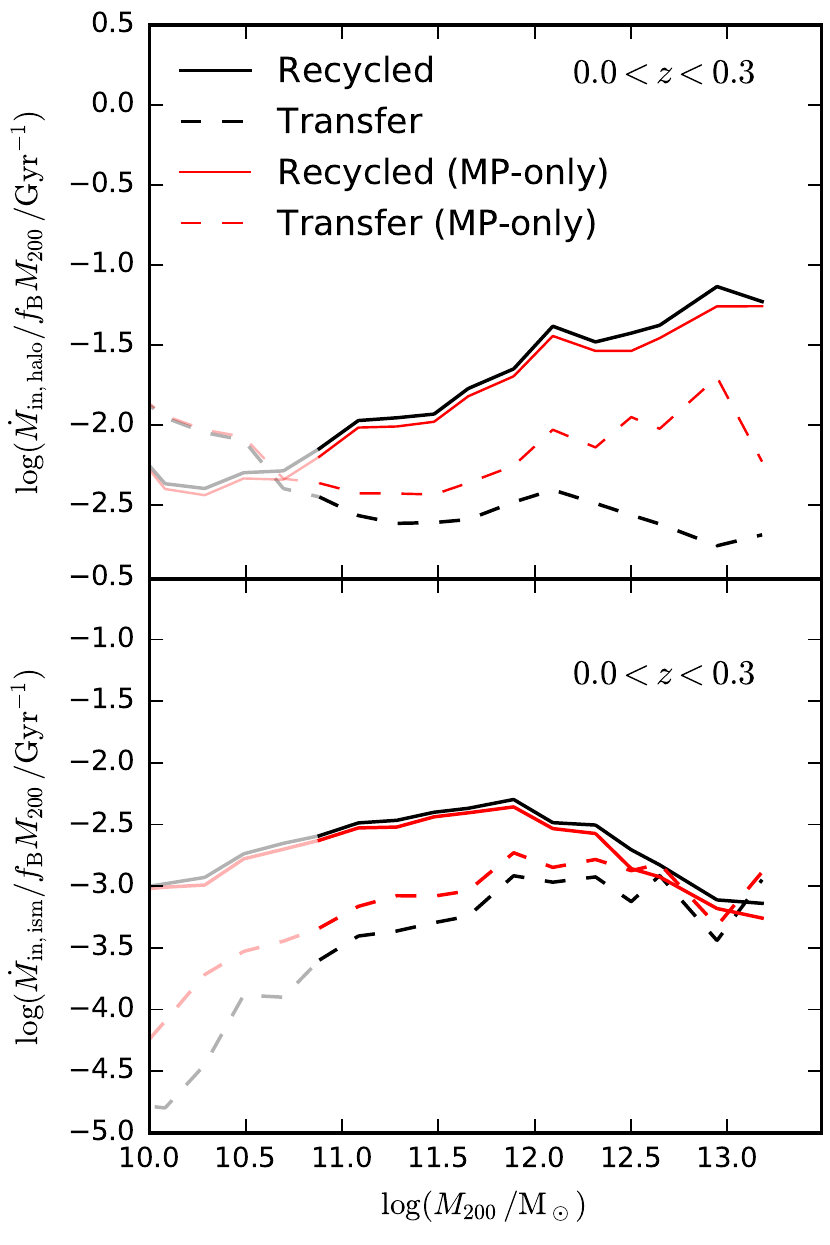}
\caption{The impact of changing the definition of recycled accretion from the fiducial definition
  (solid lines), where gas is considered recycled if it was ejected from any progenitor of the current subhalo,
  to an alternative definition (\emph{MP-only}, dashed lines) for which only gas ejected from the main progenitor
  subhalo is considered as recycled, with gas ejected from other progenitors instead labelled as
  \emph{transfer}.
  Plotted are the average rates of recycled and transferred accretion for gas being both ejected from, and then later
  accreting onto, haloes (top panel), and the ISM of galaxies (bottom panel), as a function of halo mass.
  Transparent lines indicate the range where there are fewer than $100$ stellar particles per galaxy.
  Results are plotted for redshifts $0<z<0.3$, but the conclusions are very similar at other redshifts.
  Changing the recycling definition to only gas ejected from the main progenitor makes little
  difference to our results for recycling, but increases the contribution of halo-scale gas transfer 
  at high halo masses.
}
\label{from_mp_comp_fig}
\end{figure}

In our fiducial analysis we consider ``recycled'' gas accretion as being gas that
was previously in the ISM (for our galaxy-scale measurements) of any progenitor
of the current galaxy. ``Transferred'' gas accretion is correspondingly considered as gas
that was previously in the ISM of any non-progenitor galaxy (this can include
gas that was ejected from surviving satellites of the current central galaxy).
This differs from other studies in the literature, which have defined ``recycled''
accretion as gas originating only from the main progenitor of the current
galaxy, with ``transferred'' gas in this case also including gas that was
ejected from other progenitors that have since merged with the central galaxy.

Fig.~\ref{from_mp_comp_fig} shows the effect of changing from our fiducial
definition to the latter alternative definition (which we label here as
``MP-only''). The effects of changing the definition are minor for
the recycled gas accretion rates, at both halo and galaxy scales.
The fractional impact is larger for the transferred gas accretion rates, which
are modestly increased using the alternative definition.

\subsection{Varying the velocity cut used to define recycled accretion}
\label{ap_vcut}

\begin{figure}
\includegraphics[width=20pc]{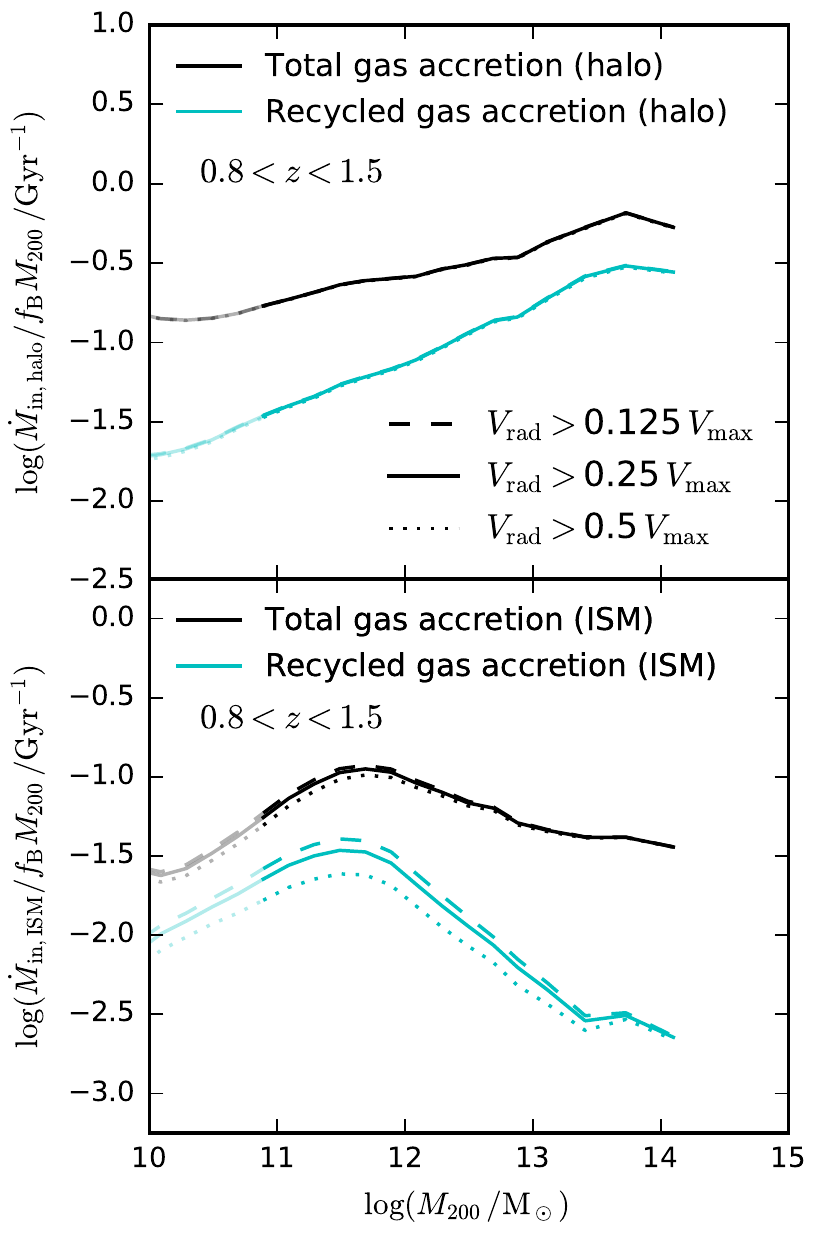}
\caption{The impact of changing the (time-integrated) radial velocity cut used to select outflowing gas that is being
  ejected from haloes (top panel) and from the ISM of galaxies (bottom panel).
  Gas accretion rates are plotted at $z \sim 1$ as a function of halo mass; results are very similar at other redshifts.
  Black (cyan) lines show total (recycled) gas accretion rates.
  Transparent lines indicate the range where there are fewer than $100$ stellar particles per galaxy.
  Changing the velocity cut by a factor two makes virtually no difference at the halo scale, 
  or for total accretion rates at the galaxy scale, 
  but has a minor effect on recycled gas accretion rates at the galaxy scale.
}
\label{vmax_cut_eff}
\end{figure}

As discussed in detail in \cite{Mitchell20}, we use a time-integrated radial
velocity cut (averaged over one quarter of a halo dynamical time) to define 
which particles have been ejected from the ISM (or halo).
This affects the amount of recycled gas accretion, with a higher velocity
cut resulting in lower recycled gas accretion rates.
Fig.~\ref{vmax_cut_eff} shows the impact of varying this cut up and down by a 
factor two from our fiducial cut (which is at $0.25 \, V_{\mathrm{max}}$, where $V_{\mathrm{max}}$
is the maximum halo circular velocity). The impact is negligible at the halo
scale, but makes a modest difference to recycling rates at the galaxy 
scale (at the level of tens of percent).

\subsection{Varying the halo mass cut used to define first-time, recycled, and transferred accretion}
\label{ap_smthr}

\begin{figure}
\includegraphics[width=20pc]{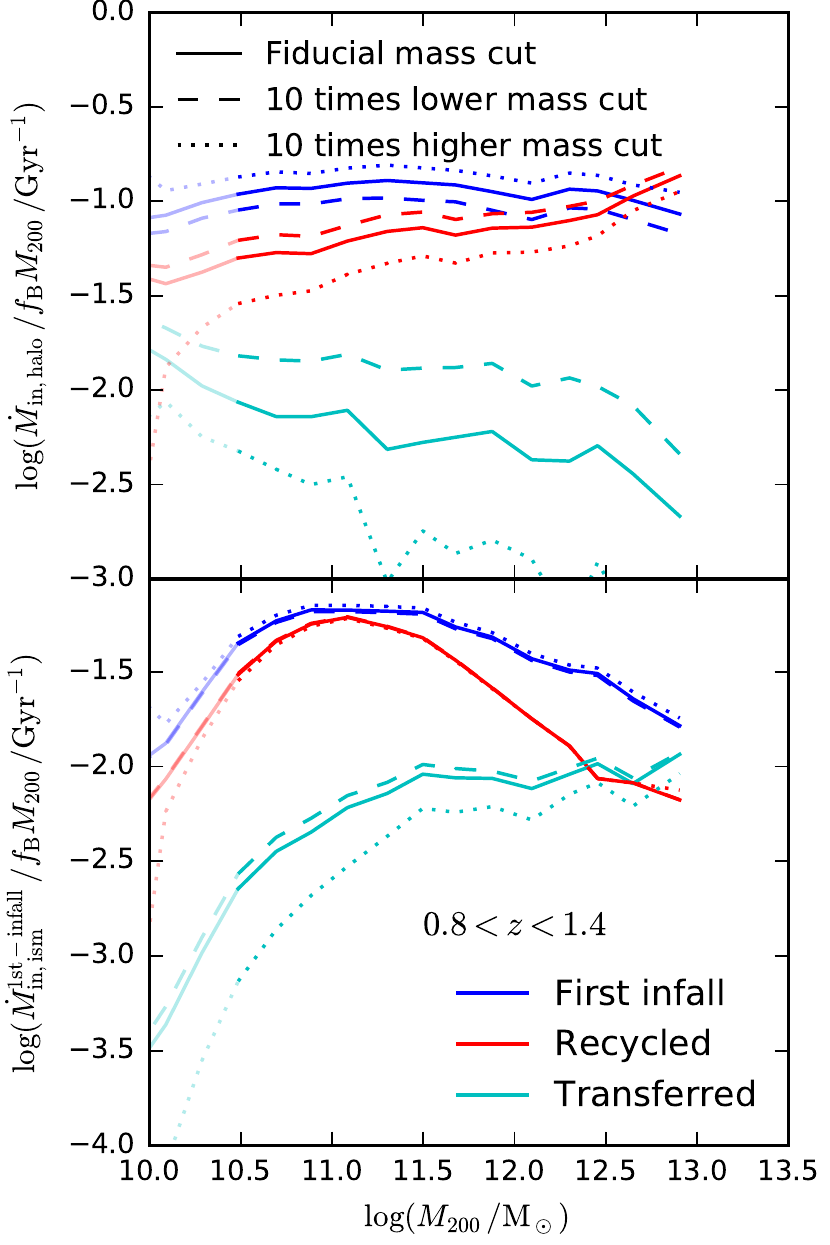}
\caption{The impact of changing the halo mass cuts used to define smooth accretion versus
  accretion of satellites, and to define the selection of first-time (blue), recycled (red), 
  and transferred (cyan) gaseous accretion (see main text for details).
  The fiducial halo mass cut (solid lines) is made at $9.7 \times 10^8 \, \mathrm{M_\odot}$,
  corresponding to $100$ dark matter particles at standard \eagle resolution.
  Here, we use the higher resolution Recal simulation to enable a meaningful exploration of changing
  our fiducial mass cut to both ten times lower (dashed) and ten times higher values (dotted).
  The top panel shows inflow rates onto haloes, and the bottom panel shows inflow rates onto the ISM of central galaxies.
  Results are plotted for $z \sim 1$, but are very similar at other redshifts.
  Transparent lines indicate the range where there are fewer than $100$ stellar particles per galaxy.
  Inflow rates are generally well converged at our fiducial mass cut (relative to the lower
  mass cut), but adopting a ten times higher mass cut would result in less recycling and transferred accretion.
  The exception is for gas being transferred onto the halo (bottom-right panel), for which our fiducial
  mass cut is not converged compared to the lower mass cut, implying that we underestimate the fraction
  of transferred accretion onto haloes in our fiducial analysis.
}
\label{m200_cut_eff}
\end{figure}

At the galaxy scale, we consider ``smooth'' accretion (as opposed to ``lumpy'' accretion via galaxy mergers) to
be gas that is accreted while not within the ISM of a subhalo with mass above a given threshold, 
which we set at $9.7 \times 10^8 \, \mathrm{M_\odot}$, corresponding to the mass of $100$ dark matter
particles at the fiducial \eagle resolution. For smooth accretion, we only consider accreted gas as being 
recycled or transferred if it was previously within the ISM of a subhalo with mass above the same threshold.
We use the same definitions at the halo scale, though in this case we base the different accretion modes
on whether gas was previously within the virial radius of a halo with mass above the threshold, without 
needing to have been within the ISM in the past.

Fig.~\ref{m200_cut_eff} shows the effect of changing this halo mass threshold up and down by a factor
ten. Generally speaking, accretion rates have converged at our fiducial mass cut (but would change
if we used a higher mass cut, dotted lines). This implies that the comparison between \eagle and 
cosmological zoom-in simulations presented in Section~\ref{acc_mode_comp_sec} is unaffected by our 
comparative inability to resolve lower-mass haloes in \eagle (due to the lower numerical resolution employed).
The exception in terms of convergence is for ``transferred'' gas at the halo
scale (cyan line, top panel), for which the rates increase by roughly a factor two if we lower the mass threshold
by a factor ten. Halo-scale gas transfer aside, we interpret the convergence of our results at a threshold
of $9.7 \times 10^8 \, \mathrm{M_\odot}$ as likely being related to the halo mass scale below which galaxies
are prevented from efficiently forming due to photo-heating from the UVB.

\label{lastpage}
\end{document}